%% file: dis-ht.final.tex
\documentclass[11pt,a4paper]{article}
\pdfoutput=1

\newlength{\dinwidth}
\newlength{\dinmargin}
\usepackage{t1enc}
\usepackage{textcomp}
\usepackage[utf8]{inputenc}
\usepackage{graphicx}
\usepackage[normalem]{ulem}

\usepackage{fancyhdr}
\usepackage{calc}
\usepackage{amssymb}
\usepackage{amsmath}
\usepackage{indentfirst}

\usepackage{pdflscape}
\usepackage{xspace}
\usepackage{color}
\definecolor{cit}{rgb}{0.,0.2,1}
\usepackage{colortbl}
\usepackage[hypertexnames,setpagesize,%
    pdftex,%
    colorlinks,%
    citecolor=cit,%
    hyperindex,%
    plainpages=false,%
    bookmarksopen,%
    bookmarksnumbered%
  ]{hyperref}
\usepackage[numbers,square,comma,sort&compress]{natbib}

\input defs.tex

\graphicspath{{figs/}}

\begin{document}

\titlepage

\begin{center}
\vspace*{2cm}
{\Large \bf  	
Evidence of quasi-partonic  higher-twist effects in deep inelastic scattering at HERA at moderate $Q^2$} \\
\vspace*{7mm}
Leszek Motyka$^{a,1}$, Mariusz Sadzikowski$^{a,2}$, Wojciech S\l{}omi\'{n}ski$^{a,3}$ and Katarzyna Wichmann$^{b,4}$
\\
\vspace*{7mm}
{\it 
$^a\;$Institute of Physics, Jagiellonian University\\
\L{}ojasiewicza 11,  30-348 Krak\'{o}w, Poland\\[1ex]
$^b$\; DESY, Notkestrasse 85, 22607 Hamburg, Germany 
} 
\\[1em]
{\it $^1$ leszek.motyka@uj.edu.pl \\
$^2$ mariusz.sadzikowski@uj.edu.pl \\
$^3$ wojtek.slominski@uj.edu.pl\\
$^4$ katarzyna.wichmann@desy.de
}\\[5mm]
 
19 July 2017 \\

\end{center}

\vspace*{2ex}

\begin{abstract}
The combined HERA data for the inclusive deep inelastic scattering (DIS) cross sections for the momentum transfer $Q^2 > 1\,\mathrm{GeV}^2$ are fitted within the Dokshitzer-Gribov-Lipatov-Altarelli-Parisi (DGLAP) framework at next-to-leading order (NLO) and next-to-next-to-leading order (NNLO) accuracy, complemented by a QCD-inspired parameterisation of twist~4 corrections. A modified form of the input parton density functions is also included, motivated by parton saturation mechanism at small Bjorken $x$ and at a low scale. These modifications lead to a significant improvement of the data description in the region of low $Q^2$. For the whole data sample, the new benchmark NNLO DGLAP fit yields $\chi^2/{\rm d.o.f.\ } \simeq 1.2$ to be compared to 1.5 resulting from the standard NNLO DGLAP fit. We discuss the results in the context of the parton saturation picture and describe the impact of the higher-twist corrections on the derived parton density functions. The resulting description of the longitudinal proton structure function $\FL$ is consistent with the HERA data. Our estimates of higher-twist contributions to the proton structure functions are comparable to the leading-twist contributions at low $Q^2 \simeq 2$~GeV$^2$ and $x \simeq 10^{-5}$. The $x$-dependence of the twist~4 corrections obtained from the best fit is consistent with the leading twist~4 quasi-partonic operators, corresponding to an exchange of four interacting gluons in the $t$-channel. 
\end{abstract}

\newpage

\input Intro.tex

\input HTmodel.tex

\input DGLAP.tex

\input Results.tex

\input Disc.tex

\section*{Acknowledgements}
We are indebted to H.~Kowalski for many interesting discussions. L.M., M.S. and W.S. gratefully acknowledge support of Polish National Science Centre (NCN) grant no.~DEC-2014/13/B/ST2/02486.

\input biblio.tex
\end{document}

%% file: defs.tex
\DeclareRobustCommand{\ie}{{\it i.e.}\xspace}

\newcommand\Eq[1]{(\ref{#1})}
\newcommand\Fig[1]{Fig.~\ref{#1}}

\newcommand\be{\begin{equation}}
\newcommand\ee{\end{equation}}
\let\vec=\boldsymbol
\newcommand\Qsat{Q _{\mathrm{sat}}}

\newcommand\pdf{f}
\DeclareRobustCommand\chidof{\ensuremath{\chi^2/\mathrm{d.o.f.}}\xspace}
\DeclareRobustCommand\FL{\ensuremath{F_{\mathrm{L}}}\xspace}
\DeclareRobustCommand\FT{\ensuremath{F_{\mathrm{T}}}\xspace}
\DeclareRobustCommand\Ftwo{\ensuremath{F_2}\xspace}

\DeclareRobustCommand\GeV{\ensuremath{\,\mathrm{GeV}}\xspace}
\DeclareRobustCommand{\alphas}{\ensuremath{\alpha_{\mathrm{s}}}\xspace}
\DeclareRobustCommand{\qsat}{\ensuremath{Q_{\mathrm{sat}}}\xspace}
\DeclareRobustCommand{\alphaem}{\ensuremath{\alpha_{\mathrm{em}}}\xspace}
\DeclareRobustCommand{\QQmin}{\ensuremath{Q^2 _{\mathrm{min}}}\xspace}
\DeclareRobustCommand{\muFstart}{\ensuremath{\mu_0}\xspace}
\DeclareRobustCommand{\muF}{\ensuremath{\mu_{\mathrm{F}}}\xspace}

\DeclareRobustCommand{\sigred}{\ensuremath{\sigma_{\mathrm{red}}}\xspace}
\DeclareRobustCommand{\uval}{\ensuremath{u_{\mathrm{val}}}\xspace}
\DeclareRobustCommand{\dval}{\ensuremath{d_{\mathrm{val}}}\xspace}

\newcommand\Ag{A_g}
\newcommand\Bg{B_g}
\newcommand\Cg{C_g}
\newcommand\Auv{A_{u-}}
\newcommand\Buv{B_{u-}}
\newcommand\Cuv{C_{u-}}
\newcommand\Euv{E_{u-}}
\newcommand\Adv{A_{d-}}

\newcommand\Cdv{C_{d-}}

\newcommand\ADbar{A_{\bar D}}
\newcommand\BDbar{B_{\bar D}}
\newcommand\CDbar{C_{\bar D}}
\newcommand\CUbar{C_{\bar U}}
\newcommand\cT{c_{\mathrm{T}}^{(0)}}
\newcommand\cL{c_{\mathrm{L}}^{(0)}}

\newcommand\cLlog{c_{\mathrm{L}}^{(\mathrm{log})}}
\newcommand\GDbar{\xds}

\newcommand{\fs}{\ensuremath{\beta}\xspace}
\newcommand{\xds}{\ensuremath{{\hat x}}\xspace}

\newcommand{\sea}{\mathrm{sea}}
\newcommand{\pds}{\ensuremath{{d}}\xspace}

\newcommand{\fitName}[1]{\mbox{\textsf{#1}}\xspace}
\newcommand{\HERAPDF}{\fitName{HERAPDF2.0}}

\definecolor{oldcolor}{rgb}{0,0,1}
\definecolor{newcolor}{rgb}{1,0,0}

\newcommand{\todo}[1]{
\underline{TODO}
\vspace*{-1.5ex}
\bgroup\quotation

#1
\egroup
}



%% file: Intro.tex
\section{Introduction and conclusions}
\label{Sec:Intro}

Good understanding of the proton structure has been one of the fundamental goals of particle physics over recent decades. Measurements of the deep inelastic $e^{\pm}p$ scattering (DIS) performed by H1 and ZEUS collaborations at the HERA collider contributed invaluable experimental input into this task. The combined data of H1 and ZEUS~\cite{Abramowicz:2015mha} that include all the measurements of the proton structure functions provide the most accurate information on the proton structure over wide range of Bjorken~$x$ and the momentum transfer $Q^2$, in particular at smaller~$x$ and $Q^2$. Hence it is crucial to fully use these data to extract the precise information on the parton density functions (PDFs) in the proton. 

The standard description of the proton structure function in QCD relies on the operator product expansion (OPE) in which only the leading---twist~2 operators---are retained. The twist~2 contributions to proton structure functions obey the hard factorisation theorem that allows to isolate the universal twist~2 parton density functions. The PDFs drive the proton scattering cross sections and the accuracy of the PDFs determination is crucial for the precision of measurements at proton colliders. It follows from the OPE however, that the twist~2 description of proton scattering is subject of higher-twist (HT) corrections that enter with suppression of inverse powers of the hard process scale squared,~$Q^2$. Those corrections, although quickly decreasing with $Q^2$, may affect the determination of the PDFs from the cross sections. In order to avoid determination error of the PDFs it is necessary to include the higher-twist terms in the analysis. Currently not much is known about higher-twist components of the proton structure. The operator content is increasingly complicated with the increasing twist and the available data are not sufficient to perform a clean and straightforward measurement of the higher-twist terms. Fortunately, the model independent characteristics of higher-twist terms given by their $Q^2$ scaling provides opportunity to obtain some information on the higher-twist corrections from fits to DIS data extended to low $Q^2$. From the theory side, properties of the leading twist~4 singularity at small~$x$ were investigated \cite{Bartels:1993it}, corresponding to a quasi-partonic \cite{BFKL2} four-gluon exchange. 
It was found \cite{Bartels:1993it} that the energy dependence of the leading exchange is up to $1/N_c^2$ corrections, given by a double gluonic ladder exchange in the $t$-channel. Hence, although the overall magnitude of the twist~4 contributions is currently undetermined, the $Q^2$ and $x$-dependencies of these contributions are known from theory and may be used as the higher-twist signatures.  Estimates of the higher-twist corrections to DIS at small~$x$ \cite{Bartels:1993ke,Martin:1998kka} and fits to the DIS data with higher-twist corrections \cite{Martin:1998np} have been performed since many years. Only recently however, with the most precise set of the combined HERA data these ideas implemented in several fits to diffractive DIS \cite{MoSadSlo} and inclusive DIS \cite{Harland-Lang:2016yfn,Abt:2016vjh} have lead to accumulating an evidence for higher-twist corrections in diffractive and inclusive DIS. 

The recent studies of higher-twist effects in inclusive DIS \cite{Harland-Lang:2016yfn,Abt:2016vjh} are based on DGLAP fits of the leading-twist contribution complemented by a simple model of twist~4 correction. In the central models elaborated in these analyses a multiplicative twist~4 correction was assumed for the longitudinal structure function $\FL$ of the form of $A \FL / Q^2$ and the twist~4 correction to the structure function $\Ftwo$ was set to vanish.\footnote{In more detail, in both the studies \cite{Harland-Lang:2016yfn,Abt:2016vjh} it was checked that a higher-twist correction to $\Ftwo$ does not improve their fits, and in Ref.\ \cite{Harland-Lang:2016yfn} some additional $x$~dependent variations of the twist correction term to $\FL$ were allowed, but they were not found to lead to a significant improvement of the result for low $Q^2$.}  The fit quality increased significantly for the combined HERA data for $Q^2 > \QQmin = 2\GeV^2$ for both NLO and NNLO DGLAP approximations of the leading-twist evolution. This simple model provides a surprisingly good description of the DIS data, except of the predicted steep rise of $\FL$ towards the low $Q^2$ for the NNLO DGLAP fit with the twist~4 correction found in Ref.\ \cite{Abt:2016vjh}. Such a rise is not inline with the $\FL$ data \cite{Collaboration:2010ry, Andreev:2013vha,Abramowicz:2014jak}. 

In this paper we adopt a more flexible model of the twist~4 contribution motivated by an extraction of the twist~4 corrections to structure functions \cite{BGBP,BGBM} from the Golec-Biernat--W\"{u}sthoff saturation model \cite{GBW}. The model is inspired by a resummation of multiple scattering in QCD in the eikonal approximation and it is capable to provide more information on the details of the higher-twist corrections and physics insight into their origin. In this approach, the twist~4 corrections to $\Ftwo$  and  $\FL$ structure functions have non-trivial $Q^2$ and $x$ dependence. In addition, we modify the standard form of the DGLAP input for the gluon and sea density, so that they are consistent with general features of parton saturation in QCD at small~$x$. With this model complemented by the NLO or NNLO DGLAP evolution of PDFs we analyse the combined HERA data on the reduced cross sections using the xFitter package \cite{Alekhin:2014irh} with suitable extensions of the code to incorporate the new features of the model. The sensitivity to the higher-twist corrections is enhanced by performing independent fits of the data sets with the momentum transfer constrained by $Q^2 > \QQmin$ and varying the limit $\QQmin$. When $\QQmin$ is larger than 20~GeV$^2$, the \chidof measures for the DGLAP fits at NLO and NNLO accuracy, with and without the higher-twist terms are close to 1.15 and exhibit a nearly flat $\QQmin$ dependence. Below $\QQmin = 20\GeV^2$,  the \chidof of the pure DGLAP fits starts growing with the decreasing values of $\QQmin$, reaching the \chidof $\simeq 1.50$ (\chidof $\simeq 1.35$) for the NNLO (NLO) accuracy at $\QQmin = 1\GeV^2$. With the higher-twist corrections included and the saturation-improved input parameterisation of the PDFs, the \chidof is only mildly increasing when $\QQmin$ decreases, and for $\QQmin =1\GeV^2$, the \chidof reaches 1.195  (\chidof $\simeq 1.215$)  for the NNLO DGLAP + HT (NLO DGLAP + HT) fit. Hence the improvement of the fit quality by adding the higher-twist corrections for $\QQmin = 1\GeV^2$ is large and particularly pronounced for the fits using the NNLO DGLAP leading-twist part. Also a good description of the HERA $\FL$ data \cite{Collaboration:2010ry, Andreev:2013vha,Abramowicz:2014jak} is obtained down to the lowest measured values of $Q^2$.   
 
The evidence for sizeable contributions of higher-twist terms is further strengthened by an explicit analysis of the twist composition of the structure functions at small~$x$ and moderate and low scales. Consistently we find the growing higher-twist effects when $x$ and $Q^2$ decrease.  The relative importance of the higher-twist corrections is found to be larger in the NNLO fit than in the NLO one. In particular, in the NNLO fit at $Q^2=1.2\GeV^2$, the twist~4 correction to the reduced cross section is found to be larger than the leading-twist contribution for $x< 2\cdot 10^{-4}$, and the relative higher-twist correction further grows towards small $x$, and at the lowest available $x \simeq 2 \cdot 10^{-5}$ it reaches about 200\% of the leading-twist term. The higher-twist effects quickly decrease with increasing $Q^2$ and reach $\sim 10\%$ level at $Q^2=6.5\GeV^2$. The higher-twist effects are found to be much stronger in the longitudinal structure function $\FL$. In particular, the twist~4 contribution to $\FL$ is larger than the leading-twist contribution for $Q^2 < 5$ GeV$^2$ ($Q^2 < 6$ GeV$^2$) for the NLO DGLAP fit (NNLO fit). Of course, also in $\FL$ the effects of higher-twist corrections decrease quickly with $Q^2$, but in the NNLO fit, the higher-twist contribution is still visible at 10\% level up to a sizeable scale $Q^2 \simeq 20\GeV^2$. 

The inclusion of higher-twist corrections is found to affect significantly the fitted gluon and sea density functions at small~$x<0.01$ and moderate factorisation scales, $\mu^2$, while the sensitivity of the valence quark distribution  to the higher-twist effects is minor and may be neglected. The largest difference in PDFs coming from the higher-twist effects is found in the gluon PDF---the difference is large and much larger than the corresponding uncertainty at small scales, 1~GeV$^2 < \mu^2 < 3.5$~GeV$^2$, then decreasing with $\mu^2$ to a few percent level at $\mu^2 = 50$~GeV$^2$. The sea distribution at small~$x$ is also affected but it exhibits lower sensitivity to the presence of higher-twist corrections.

In conclusion, we have found a consistent evidence of the sizeable twist~4 corrections to proton structure functions. The evidence comes primarily from the $\chi^2$  quality measure of fits to the combined HERA data on the inclusive DIS with the leading-twist component described by the NLO / NNLO DGLAP evolution. This evidence is further strengthened by the strong effects of the higher-twist corrections in the reduced cross section and the structure function $\FL$ at small~$x$ and moderate / low $Q^2$. The fitted twist~4 contributions have the $x$-dependence that is consistent with the double exchange of hard gluonic ladder at small~$x$, as expected from the QCD analysis of the evolution of leading quasi-partonic operators \cite{Bartels:1993it,BFKL2}.


%% file: HTmodel.tex
\section{The model of higher-twist corrections}
\label{Sec:2}


The Golec-Biernat--W\"{u}sthoff (GBW) saturation model~\cite{GBW} offers a simple and effective description of DIS, DDIS structure functions down to very low $Q^2$ at small $x$, and also of the exclusive vector meson production~\cite{Kowalski:2006hc}. In particular, with this model one is able to describe reasonably well even the transition from DIS at large $Q^2$ to the photoproduction limit. This transition may be viewed as a transition from the twist~2 regime to the region that all twist contributions are relevant. From the point of view of perturbative QCD, the GBW model corresponds to multiple independent high-energy scatterings of photon hadronic fluctuations, that is to the eikonal iteration of a single gluon ladder exchange. In particular, the leading behaviour of twist~4 contributions of the GBW cross section is $\sim (x^{\lambda}Q^2)^{-2}$ (modulo logarithms) compared to the leading twist cross section $\sim (x^{\lambda}Q^2)^{-1}$ (modulo logarithms) \cite{BGBP,BGBM}. Such behaviour of twist~2 and twist~4 amplitudes is in a qualitative agreement with results of the evolution of twist~4 contributions in the Bukhvostov, Frolov, Lipatov and Kuraev framework \cite{BFKL2,Bartels:1993it,BraunTwist}, where dominant contributions at small~$x$ are driven by quasi-partonic operators.


From these studies \cite{BGBP,BGBM} it follows that the twist~4 contributions to the transverse and longitudinal cross sections take the form
\begin{equation}
\label{sigma_GBW}
\sigma^{(\tau=4)}_{\mathrm T} = A\left[\frac{\qsat^2(x)}{Q^2}\right]^2,\;\;\;
\sigma^{(\tau=4)}_{\mathrm L} = -\frac{4}{3}A\left[\frac{\qsat^2(x)}{Q^2}\right]^2 \left[\log\left(\frac{Q^2}{\qsat^2(x)}\right)+B\right],
\end{equation}
where $A, B$ are positive constants. The saturation scale depends on $x$ variable as $\qsat^2= Q_0^2 (x_0/x)^{\lambda}$
where $Q_0^2 = 1$ GeV$^2$ and $x_0 = 3.04\cdot 10^{-4}$ (the GBW fit without charm) are model parameters. It
is important to notice the opposite signs of the corrections in (\ref{sigma_GBW}), positive for the transverse and negative for the longitudinal part.
Additionally, the twist~4 contribution to the longitudinal cross section is logarithmically enhanced. This structure of the corrections can be also deduced from the general QCD analysis which was discussed in \cite{BGBP}. Following the above considerations we shape a saturation model inspired ansatz for twist~4 corrections in the following form
\begin{equation}
\label{twist_ansatz}
  \Delta\sigma_{\mathrm{L/T}} ^{(\tau = 4)} = \left[\frac{\qsat^2(x)}
  {Q^2}\right]^2 \left[
  \tilde c_{\mathrm{L/T}}^{\mathrm{(log)}}
  \log\left(\frac{\qsat^2(x)}{Q^2}\right) +\tilde c_{\mathrm{L/T}} ^{(0)}\right],
\end{equation}
in which coefficients $\tilde c_{\mathrm{L/T}} ^{(0)}$ and $\tilde c_{\mathrm{L/T}} ^{\mathrm{(log)}}$ are left as free and independent parameters. In practical implementation of the fits we rewrite the above parameterisation to the following convenient form of twist~4 corrections to the structure functions:
\begin{equation}
\label{delta4F}
\Delta F^{(\tau=4)}_{\mathrm{L/T}} = \frac{Q_0 ^2}{Q^2} x^{-2\lambda} \left[ c_{\mathrm{L/T}} ^{\mathrm{(log)}} \left(\log\frac{Q_0^2}{Q^2} + \lambda \log \frac{1}{x}\right) + c_{\mathrm{L/T}} ^{(0)} \right],
\end{equation}
where $F_{\mathrm {L/T}} = Q^2\sigma_{\mathrm{L/T}} / 4\pi^2\alphaem$.

Equation (\ref{twist_ansatz}) contains the $\tilde c^{\mathrm{(log)}} _{\mathrm T}$ parameter which determines the magnitude of the logarithmically enhanced term in the transverse cross section. Due to the properties of the transverse photon impact factor, this parameter vanishes at the leading order in the strong coupling constant expansion \cite{Bartels:1994jj,BGBP,BGBM}. Therefore, one expects that this parameter is much smaller than parameter $\tilde c_{\mathrm T} ^{(0)}$, hence it is neglected in our fits---we assume $\tilde c_{\mathrm T}^{\mathrm{(log)}} \equiv c_{\mathrm T}^{\mathrm{(log)}} = 0$\footnote{We confirmed that this parameter is indeed small in independent numerical fits.}.

When one attempts to extend the proton structure function analysis to the region of small $Q^2$ and very small $x$ it is necessary to consider possible effects of parton rescattering and/or recombination in the dense parton regime. Those effects may lead not only to the higher-twist corrections but also they are expected to influence the form of the input for QCD evolution of the matrix elements in the OPE. In particular, it is natural to require that the input functions for the DGLAP evolution of parton densities are consistent with unitarity constraints relevant for high-energy scattering at very low~$x$. Hence, a precise analysis of the higher-twist effects which pronounce at low $Q^2$ and small~$x$ requires careful treatment of the gluon and sea input distributions at small~$x$.

Currently, one of the most successful tools for an analysis of $\gamma^*$--nucleon (or $\gamma^*$--nucleus) scattering with the unitarity corrections is the Balitsky--Kovchegov (BK) equation \cite{BK:Bal,BK:Kov}. In constructing the model for initial conditions of the PDFs we take into account the outcome of this equation analysis. The BK equation, that resums multiple scattering effects in the extended generalised leading logarithmic $1/x$ approximation \cite{Bartels:1992ym,EGGLA} and the large $N_c$~limit, may be described in a natural way in terms of the colour dipole language \cite{NZ,Mueller:1993rr,BK:Kov}. In this approach the high-energy scattering is described in terms of the imaginary part of the BK dipole forward scattering amplitude, $N(x,\vec{r})$, where $\vec{r}$ is the dipole extension vector in the transverse plane. On the other hand the same dynamics may be covered by the BK equation represented in terms of the unintegrated gluon density \cite{Kutak:2003bd,Kutak:2004ym,Bartels:2006ea,Bondarenko:2006ft}. It can be shown that at the leading logarithmic $1/x$ accuracy the imaginary part of the BK dipole forward scattering amplitude, $N(x,\vec{r})$, and the BK unintegrated gluon density\footnote{Note that we use for the unintegrated gluon density the normalisation convention that is defined by a relation to the collinear gluon density $f_g(x,\mu^2)$: $xf_g(x,\mu^2) = \int ^{\mu^2} d\vec{k}^2 {\cal F}(x,\vec{k}^2)$.} ${\cal F}(x,\vec{k}^2)$ (where $\vec{k}$ is the gluon transverse momentum) are in one-to-one correspondence. Hence the unintegrated gluon density at small~$x$ can be recovered from $N(x,\vec{r})$,
\begin{equation}
 {\cal F}(x,\vec{k}^2) = \frac{N_c R_p^2}{2\alpha_s\pi}\vec{k}^2\nabla_{\vec{k}}^2 \tilde N(x,\vec{k}),
\end{equation}
where $\tilde N(x,\vec{k}) = \int d^2\vec{r} \exp(-i\vec{k}\vec{r}) \, N(x,\vec{r})$ is the Fourier transform of the dipole scattering amplitude, and $R_p$ is an effective radius of the proton.
After employing the leading logarithmic relation of the collinear gluon density $f_g(x,\mu^2)$  and the unintegrated gluon density
${\cal F}(x,\vec{k}^2)$ one obtains
\begin{equation}
xf_g(x,\mu^2) = \frac{N_c R_p^2}{2\alpha_s\pi}\int^{\mu^2}d\vec{k}^2\, \vec{k}^2\,  \nabla_{\vec{k}}^2 \tilde N(x,\vec{k}).
\end{equation}
Explicit numerical solutions of the BK equation \cite{GBMS,Bondarenko:2006ft} show that at small $x$ and for $\vec{k}$ below the saturation scale, $\qsat(x)$, generated by the BK evolution, the solution of the BK equation tends to ${\cal F}(x,\vec{k}^2) \sim R_p ^2 \vec{k}^2 / \qsat^2(x) \sim x^{\lambda}$. Exactly the same asymptotic behaviour of the unintegrated gluon density is found in the GBW model where one approximates the scattering amplitude with the saturation formula $N(x,\vec{r})\sim 1-\exp(-\vec{r}^2 \qsat^2(x))$. In fact, the small $\vec{k}$ asymptotics of the saturated unintegrated gluon density may be traced back to the unitarity constraint on the dipole cross section in the position space, $N(x,\vec{r}) \leq 1$. For such a form of  ${\cal F}(x,\vec{k}^2)$ at small $\vec{k}$, it is straightforward to show that in the limit $\mu^2\ll \qsat^2(x)$ the gluon density small~$x$ asymptotics is
\begin{equation}
xf_g(x,\mu^2) \sim x^\lambda,
\end{equation}
with $\lambda > 0$. Therefore the gluon density at a low scale is expected to decrease toward zero with decreasing $x$.
Following this argument our parameterisation of the input for gluon distribution fulfils the condition $xf_g(x,\mu_0^2) \sim x^{B_g}$, at $x\to 0$, where $B_g$ is a positive fit parameter.

Furthermore in our model construction we consider the input for the sea distribution at a small scale and small $x$. Assuming that the sea quarks at small $x$ are generated predominantly from the gluon DGLAP splitting to quark/antiquark, one may approximate the sea singlet distribution by the LO DGLAP expression describing the feed-down from gluons,
\begin{equation}
f_{\mathrm{sea}}(x,\mu_0^2) \simeq \int ^{\mu_0 ^2} \frac{d\mu^2}{\mu^2} \,\frac{\alphas(\mu^2)}{\pi} \int_x ^1 \frac{dz}{z} P_{qg}(z)f_g(x/z,\mu^2).
\label{eq:seainput}
\end{equation}
Obviously, since the scales probed are low and the impact of non-perturbative effects unknown, the above expression should be treated only as a QCD hint on the actual shape of the sea distribution at low scales. An explicit evaluation of the model expression in Eq.\ (\ref{eq:seainput}) leads to the sea-quark distribution asymptotic behaviour at small~$x$ following the gluon asymptotics, $xf_{\mathrm{sea}}(x,\mu^2) \sim x^\lambda$. Hence in the fits that include parton saturation effects in the input distributions, we impose the same asymptotic behaviour of the sea and gluon distributions at the initial scale,  $xf_{\mathrm{sea}}(x,\mu_0^2) \sim xf_g(x,\mu_0^2) \sim x^{B_g}$.


%% file: DGLAP.tex
\section{DGLAP framework}
\label{Sec:DGLAP}

The leading twist~2 contributions to the \Ftwo and \FL structure functions are given in terms of PDFs, $\pdf_k (x,\mu^2)$, determined within the DGLAP framework. Our approach follows closely the scheme adopted in the \HERAPDF~\cite{Abramowicz:2015mha} study, in order to clearly see the effects of higher-twist contributions.
The PDFs are parameterised at the starting scale $\mu_{\mathrm{F}0}$
and then determined at all scales $\muF$ by solving the DGLAP evolution equations. 
The factorisation and renormalisation scales are chosen to be equal
and in the following we denote them by $\mu$,
while the evolution starting scale 
is denoted by \muFstart.

\subsection{Scheme description}
\label{Sec:DGLAP.scheme}

The light quarks, $u,d,s$, are taken to be massless.
The heavy quarks, $c,b,t$, are generated radiatively and appear only at transition scales, taken to be equal to the corresponding quark masses, $m_h$.
The PDFs of heavy quarks start from 0 once $\mu$ goes above $m_h$. 
In other words, there are no intrinsic heavy flavours.
For a simple realisation of this scenario we take the starting scale, \muFstart, below the charm mass.

The coefficient functions of heavy quarks are calculated in the Thorne--Roberts general-mass
variable-flavour-number scheme called RT OPT \cite{Thorne:1997ga,Thorne:2006qt,Thorne:2012az}.
This scheme is adopted in accordance with the \HERAPDF fit \cite{Abramowicz:2015mha}.


\subsection{The input parameterisation}
\label{Sec:DGLAP.input}

The distributions parameterised at the starting scale include the gluon $g$, $d$ and $u$ valence quarks,
and up- and down-type sea quarks, $\bar U = \bar u, \bar D = \bar d + \bar s$.

The generic form of the input parton $k$~distribution, $x\pdf_k(x) = x\pdf_k(x,\mu_0^2)$ is assumed to be,
\begin{equation}
\label{eq:pdf.para.gen}
x \pdf_k(x) =
A_k x^{B_k} (1-x)^{C_k} \left(
  1 + {D_k} x + {E_k} x^2
\right)
\,,
\end{equation}
for $k = g, \uval, \dval, \bar{U}, \bar{D}$.

The relative $\bar{s}$ contribution to the down-type sea at the starting scale is
assumed to be a fixed ($x$-independent) fraction \fs of $\bar{D}$,
\ie $f_{\bar{s}} = \fs  f_{\bar{D}}$.

Thus, in general, we have 26 fit parameters to start with.
Several assumptions are made in order to make this number smaller.

First, the quark-counting and momentum sum rules are used to fix the valence and gluon PDFs normalisation parameters,
$A_{\uval}, A_{\dval}$  and $A_g$.
Next, the \uval and \dval distributions are assumed to have the same shape at small $x$ \ie
$B_{\uval} = B_{\dval}$\footnote{This constraint is not used in the \HERAPDF fit.}.
Also, to ensure a uniform sea behaviour at low $x$, ($\bar{u} \simeq \bar{d}$),
the following constraints are imposed:
$A_{\bar{U}} = (1-\fs) A_{\bar{D}}$
and
$B_{\bar{U}} = B_{\bar{D}} \equiv B_\sea$.
Finally, the strange sea fraction, \fs, is set to 0.4.
Based on the \HERAPDF experience and our numerous fit results we set to zero all $D_k$ and $E_k$, except of $E_{\uval}$ which is left free. With this setup we have 10 free parameters of the input PDFs to be compared with 14~free parameters of the \HERAPDF fit. With this restricted parameterisation at the leading-twist level,
we gain on the stability of the fits with the saturation and higher-twist effects included.

The first step towards improved description at  low $x$ and low $\mu^2$ is a modification of the basic parameterisations \Eq{eq:pdf.para.gen}, aimed at improving description of parton saturation effects in the gluon and sea input distributions in the small~$x$ domain.

In the \HERAPDF fits the input gluon parameterisation
is augmented by a negative term
$-A'_g x^{B'_g} (1-x)^{25}$ \cite{Abramowicz:2015mha,Martin:2009iq}. In the current study we do not include this subtraction.
Instead, we consider enhancing the basic parameterisation \Eq{eq:pdf.para.gen} with saturation-inspired, damping factors for the parton $k$:
\begin{equation}
\label{eq:dmp}
\left[ 1+\displaystyle\left(\frac{\xds_k}{x}\right)^{\pds_k}\right]^{-1} ,
\end{equation}
applied to the gluon and sea components, with $\pds_k+ B_k > 0$. The application of such factors ensures a smooth decrease to zero of  $xf_g(x,\muFstart^2)$ and $xf_\sea(x,\muFstart^2)$ when $x \to 0$, consistent with the known results from analyses of  parton saturation at small~$x$ (see Sec.\ \ref{Sec:2} for a more detailed discussion). The damping factors describing the parton saturation effects turn on for~$x$ below a specific scale $\xds_k$, which can be therefore interpreted as the saturation $x$ at $Q_0$: $\Qsat(\xds_k) = Q_0$. In general, $\xds_k$ and the saturation powers $\pds_g, \pds_\mathrm{sea}$ are arbitrary parameters, with the already mentioned constraint $\pds_k + B_k > 0$.

Hence we consider the following input parameterisations of the gluon and sea PDFs:
\begin{equation}
xf_k(x,Q_0^2) = A_k x^{B_k} (1-x)^{C_k} \left[ 1+\displaystyle\left(\frac{\hat x_k}{x}\right)^{d_k}\right]^{-1},
\end{equation}
where \(k=g,\bar{U}, \bar{D}\).
With these parameterisations in the \(x \to 0\) limit the input PDFs scale as
\begin{equation}
xf_k(x,Q_0^2) \simeq A_k x^{B_k+d_k}.
\end{equation}

After a preliminary analysis of the data we found that for the gluon input distribution the saturation damping factor is irrelevant, as the fits yielded \(B_g \sim 1\) which already guarantees power-like approach to zero of $xf_g(x,Q_0^2)$ for $x\to 0$.
Thus we retain the damping factor for the quark sea only, with \(\hat x_D = \hat x_U \equiv \hat x\) being a free fit parameter. For the saturation damping exponent for the quark sea, $d_{\sea}$, we impose an additional constraint following from the assumption that the sea input distribution at small $x$ follows the power-like behaviour of the gluon input distribution, see Sec.\ \ref{Sec:2}. As a result, a relation of the exponents is obtained: $B_{\sea} + d_{\sea} = B_g$, resulting in $d_{\sea} = B_g - B_{\sea}$. In fact, we have checked that leaving $d_{\sea}$ as a free parameter does not improve the fit quality (the difference in the $\chi^2$/d.o.f.\ is smaller than 0.002). Hence the phenomenological inclusion of the saturation effects in the input of the PDFs is reduced to taking the positive definite gluon input and imposing the sea input damping at small~$x$.


%% file: Results.tex
\section{Results} 
\label{Sec:Results}



In the current analysis we use the combined HERA data on neutral and charged current $e^+ p$ and $e^- p$ inclusive cross
sections, measured at centre-of-mass energies ranging from 225~GeV to 318~GeV \cite{Abramowicz:2015mha}.

In the fits we use only data points for which $Q^2 > 1\GeV^2$.
Their kinematic range spans four orders of magnitude in $x$ and $Q^2$
with lower bounds at $x = 1.76\cdot 10^{-5}$ and $Q^2 = 1.2\GeV^2$.
The inelasticity $y$ values are between 0.001 and 0.95.
The whole data set comprises 1213 data points.
A subset of this data set with $Q^2 \geq 3.5$~GeV$^2$ was used to extract the \HERAPDF PDFs~\cite{Abramowicz:2015mha}.

The measured cross sections are presented\footnote{Data set 1506.06042 at http://xfitter.hepforge.org/data.html}
in terms of the reduced cross section:
\begin{equation}
\sigred(x,Q^2,y) =
F_2(x,Q^2) - \frac{y^2}{1+(1-y)^2} \,\FL(x,Q^2)
\equiv
\FT(x,Q^2) + \frac{2(1-y)}{1+(1-y)^2} \,\FL(x,Q^2)
\,,
\end{equation}
where $\Ftwo(x,Q^2) = \FT(x,Q^2) + \FL(x,Q^2)$.  

In order to fit the data down to $Q^2 = 1.2\GeV^2$, we take the starting scale for the DGLAP evolution
$\muFstart^2 = 1\GeV^2$. For the sake of comparison to the \HERAPDF fits we present also some results for the PDFs parameterised at  $\muFstart^2 = 1.9\GeV^2$. 

The fits are performed using the xFitter package \cite{Alekhin:2014irh} supplemented by us
with necessary code extensions including input parameterisation with saturation damping effects,
and the higher-twist contributions to \FT and \FL, as given by Eq.\  \ref{delta4F}. 
The DGLAP evolution is performed using the QCDNUM program \cite{Botje:2010ay}.

\noindent In the analysis below we use the following fit names:
\begin{itemize}
\setlength\itemsep{0.1em}
\item \fitName{HTS} ---
  Higher Twist + Saturation, $\muFstart^2=1.0\GeV^2$,  $\QQmin = 1.2\GeV^2$
  \\
  --- Our main fit. It corresponds to the standard DGLAP evolution of the leading twist terms, in which the saturation damping effects are assumed in the input PDFs, complemented by the additive twist~4 corrections. Note that when performing the $\chi^2$ scan (see Sec.\ \ref{Sec:chi-scan}) we use the same name for this model fitted to data with a variable lower cutoff  $\QQmin$ on $Q^2$;
\item \fitName{HT} ---
  Higher Twist without Saturation, $\muFstart^2=1.0\GeV^2$,  $\QQmin = 1.2\GeV^2$\\
   --- Like \fitName{HTS} but without the saturation damping effects in the initial PDFs;
\item \fitName{LT-STD} ---
  Standard Leading Twist without Saturation, $\muFstart^2=1.9\GeV^2$,  $\QQmin = 3.5\GeV^2$
  \\
  --- \HERAPDF-like fit --- pure DGLAP approach, no saturation damping in the input parameterisation, the initial scale and the data selection as in  \HERAPDF;
\item \fitName{LT-1.0} ---
  Leading Twist, without Saturation, $\muFstart^2=1.0\GeV^2$ \\
  --- Like \fitName{LT-STD}, but with lower initial scale and a variable lower cutoff $\QQmin$ on $Q^2$;
\item \fitName{LT-1.9} ---
  Leading Twist, without Saturation, $\muFstart^2=1.9\GeV^2$ \\
  ---  Like \fitName{LT-STD}, but with a variable lower cutoff $\QQmin$ on $Q^2$.
\end{itemize}

\newdimen\szer

\begin{figure}[htb]
\centerline{\includegraphics[width=0.9\columnwidth]{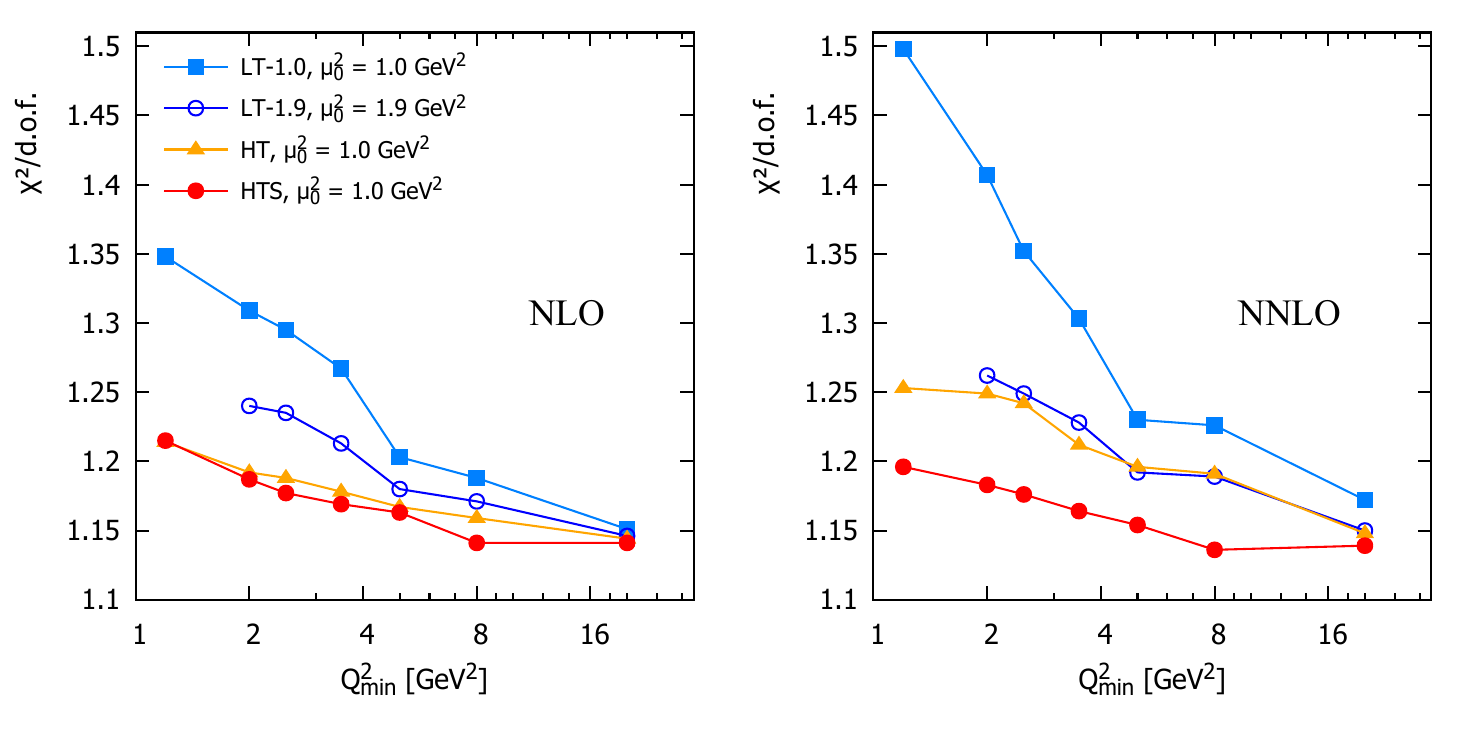}}
\vspace*{-3ex}
\caption{The \chidof of various fits to the data with the $Q^2 \ge \QQmin$ condition.}
\label{fig:chi2vsQmin}
\end{figure}

\begin{figure}[htb]
\centerline{\includegraphics[width=0.9\columnwidth]{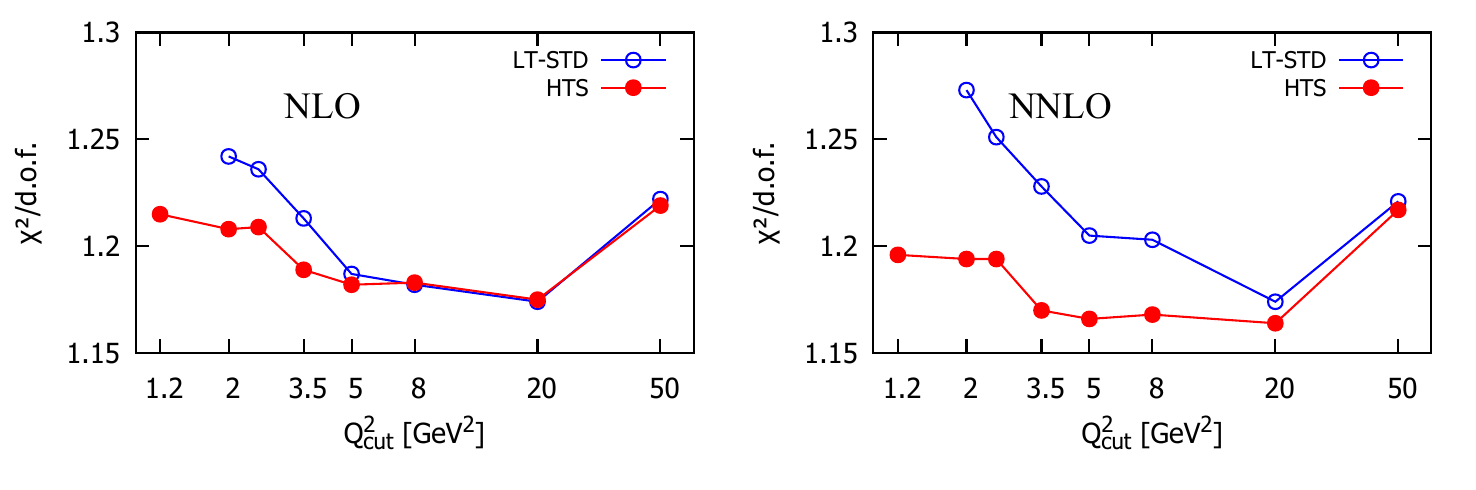}}
\vspace*{-3ex}
\caption{The \chidof for the \fitName{LT-STD} and \fitName{HTS} fits vs. lower $Q^2$ cut, $Q_\mathrm{cut}^2$, of the selected data subsamples.}
\label{fig:chi2vsQcut}
\end{figure}

The analysis of the data is carried out in the following stages. First, we perform the scan of the $\chi^2$/d.o.f.\ for the data set with  $Q^2 > Q^2_{\mathrm{min}}$ as a function of $Q^2_{\mathrm{min}}$ for all chosen setups, that is for NLO and NNLO DGLAP evolution of the leading twist, with and without the higher-twist corrections, with and without saturation damping effects in the input for the PDFs. 
We have checked that the saturation damping modification of the input distributions becomes important only upon the inclusion of the twist~4 contributions. Without the higher-twist corrections the damping gives practically no improvement of the fits, so we do not present results for the DGLAP fits without the higher-twist corrections with the saturation damping of the input. Next we describe the features of our best fits with the higher-twist corrections. Furthermore we explicitly study the effects of the higher twists for $\sigma_{\mathrm{red}}$ and $\FL$, as given by the best fits. Finally we show the impact of the inclusion of the higher-twist and saturation effects on the obtained PDFs.

\vspace*{-1ex}

\szer=0.9\columnwidth

\subsection{The \texorpdfstring{$\chi^2$}{chi squared} scans}
\label{Sec:chi-scan}

In Fig.\ \ref{fig:chi2vsQmin} we show the $\chi^2$/d.o.f.\ for the fits to the reduced cross sections as a function of the lower cutoff $Q^2_{\mathrm{min}}$ imposed on the photon virtuality $Q^2$ for the data sample taken into account in the fits.  The initial scale of the DGLAP evolution is set to \muFstart. The higher-twist parameterisation provides the best description of the data from $Q^2_{\mathrm{min}}\simeq 16\GeV^2$ below. The question of the parton saturation effects in the input sea distribution is more subtle. From the $\chi^2$ scan it follows that within the NLO approximation this is not an important effect. However, in the fits assuming the NNLO DGLAP evolution of the leading twist contribution, both the higher-twist corrections and the parton saturation effects in the input are key ingredients for the best description of the data. The inclusion of both effects  improves the data description significantly for  $Q^2_{\mathrm{min}} < 20\GeV^2$.

In Fig.\ \ref{fig:chi2vsQcut} we show a comparison of the $\chi^2/$d.o.f.\ between the data and two fits, standard fit \fitName{LT-STD} and the best fit with higher-twist corrections \fitName{HTS}, as a function of the lower cutoff $Q^2_{\mathrm{cut}}$ imposed on the data. Here $Q^2_{\mathrm{min}}$ is kept fixed to 2.0~GeV$^2$ for the \fitName{LT-STD} fit and to 1.2~GeV$^2$ for the \fitName{HTS} fit. A systematic improvement of the data description is clearly seen for the \fitName{HTS} fit with respect to the \fitName{LT-STD} fit for  $Q^2_{\mathrm{cut}} < 5$~GeV$^2$ at NLO and for  $Q^2_{\mathrm{cut}} < 20$~GeV$^2$ at NNLO.

A short comparison of our reference fit \fitName{LT-STD} to the \HERAPDF fit is in order. 
The resulting values of the $\chi^2/$d.o.f.\ for the \fitName{LT-STD} fit are 1.213 and 1.228 at NLO and NNLO, correspondingly, while for the \HERAPDF fit they are 1.200 and 1.205 at NLO and NNLO, correspondingly.
Recall however, that the input PDFs parameterisation of the \fitName{LT-STD} fit has 10 free parameters, to be compared with 14 parameters of the \HERAPDF fit, and thus the slight difference in the $\chi^2/$d.o.f.\ is acceptable.

\subsection{Features of the best fits}
\label{Sec:HTSfit}

The parameters of the model obtained from the best fits (\fitName{HTS}) of all the data with $Q^2 >1\GeV^2$ with the twist~4 corrections included and the sea input with saturation damping, are given in Tab.~\ref{tab:1}. The results are displayed both for NLO and NNLO DGLAP evolution of the leading twist terms.

\begin{table}[htbp]
\renewcommand{\arraystretch}{1.2}%
\begin{center}
\begin{tabular}{lll}
 & NLO & NNLO \cr
\hline
\\[-2ex]
\(\Ag\)     & 23.1 & 33.44 \cr
\(\Bg\)     & 0.649 ± 0.074 & 0.828 ± 0.071 \cr
\(\Cg\)     & 11.1 ± 1 & 11.0 ± 1.1 \cr
\(\Auv\)    & 5.584 & 6.171 \cr
\(\Buv\)    & 0.830 ± 0.029 & 0.870 ± 0.032 \cr
\(\Cuv\)    & 4.509 ± 0.087 & 4.550 ± 0.086 \cr
\(\Euv\)    & 8.8 ± 1.3 & 8.9 ± 1.3 \cr
\(\Adv\)    & 3.238 & 3.623 \cr
\(\Cdv\)    & 3.87 ± 0.19 & 3.93 ± 0.19 \cr
\(\ADbar\)  & 0.285 ± 0.027 & 0.233 ± 0.022 \cr
\(\BDbar\)  & -0.030 ± 0.017 & -0.077 ± 0.017 \cr
\(\CDbar\)  & 5.28 ± 0.92 & 4.56 ± 0.87 \cr
\(\CUbar\)  & 3.96 ± 0.55 & 3.86 ± 0.63 \cr
\(\lambda\) & 0.351 ± 0.014 & 0.257 ± 0.016 \cr
\(\GDbar\times 10^5\)  & 0.09 ± 0.55 & 20.0 ± 3.8 \cr
\(\cT\times 10^5\)     & 12.6 ± 4.9 & 229 ± 80 \cr
\(\cL\times 10^5\)     & 160 ± 38 & 821 ± 186 \cr
\(\cLlog\times 10^5\)  & -43 ± 12 & -326 ± 95 \cr
\chidof & 1.215 & 1.195 \cr
\end{tabular}
\end{center}
\caption{Parameters and the \chidof values for the \fitName{HTS} NLO and NNLO fits. Note that \(\Ag\), \(\Auv\) and \(\Adv\) are fixed by the sum rules.
\label{tab:1}}
\end{table}

It is interesting to analyse the obtained parameters describing the higher-twist corrections and compare the results to the expectations from the GBW model. First of all, the value obtained from the fit of the saturation scale exponent is $\lambda =  0.351 \pm 0.008$ for the NLO \fitName{HTS} fit and $\lambda =  0.257 \pm 0.016$ in the NNLO \fitName{HTS} fit. These values are rather close to the saturation exponents of the GBW model $\lambda = 0.288$ (without charm) and $\lambda = 0.277$ (with charm). The obtained values of $\lambda \sim 0.26 - 0.35$ are also consistent with the picture of double hard pomeron exchange as the leading contribution to twist~4 corrections at small~$x$. Also the value of the saturation $x$ parameter at $Q_0 =1\GeV$ obtained from the NNLO fit, $\xds = (2.0 \pm 0.4) \cdot 10^{-4}$  compares well to the corresponding GBW saturation $x$ parameters, $x_0 =  3.04 \cdot 10^{-4}$ (without charm) and  $x_0 = 0.41 \cdot 10^{-4}$ with charm. In the NLO fit, however, the obtained $\xds =  0.09 \pm 0.2 \cdot  10^{-5}$ is consistent with zero. Recall that in our approach parameter $\xds$ is the characteristic $x$ for emergence of the saturation damping effects in the sea distribution. Hence the conclusion implied by the $\chi^2$ scan is confirmed: that the sea saturation damping is important for the NNLO DGLAP fit with twist 4 corrections, while the DGLAP NLO fit with higher twists does not require the saturation input damping in the $x$~range of the fitted data. 

Interestingly, the pattern of the twist~4 multiplicative coefficients is found to differ significantly from the predictions of the GBW model. At small~$x$ the model yields a sizeable negative twist~4 correction to \FL and a positive correction to $\FT$. The performed fits exhibit a different pattern --- both at NLO and NNLO we find a small positive twist~4 correction to $\FT$ and a larger positive correction to $\FL$. The difference of the sign of the higher twist correction to $\FL$ at small~$x$ between the GBW model prediction and the fit results, occurs both in the leading logarithmic term $\propto c_{\mathrm L} ^{\mathrm{(log)}}\log(Q^2 / \qsat ^2 (x))$ for $Q > \qsat(x)$ and in the constant term $c_{\mathrm L}^{(0)}$. This result indicates that the leading twist~4 coefficient function in the longitudinal virtual photon inelastic scattering off the proton is not of the type of an eikonal pomeron exchange.

\subsection{Comparison with the data for \texorpdfstring\sigred{sigma\_reduced} and \texorpdfstring\FL{F\_L}}

In the approach presented here  the relative importance of the higher-twist corrections to the proton structure functions may be estimated for different $x$ and $Q^2$. Such an estimate provides a measure of both the expected accuracy of the leading twist description and the sensitivity to the higher-twist contribution. It also permits to determine the kinematic region in which the higher-twist corrections are most important and, in this way, the evidence of the higher-twist contribution to the structure functions is strengthened. Indeed, we find that the most important effect in the structure function deviations from the DGLAP leading twist description comes from the region of small~$x$ and $Q^2$, where the higher-twist effects are strongest.

\begin{figure}[h]
\centerline{\includegraphics[width=\szer]{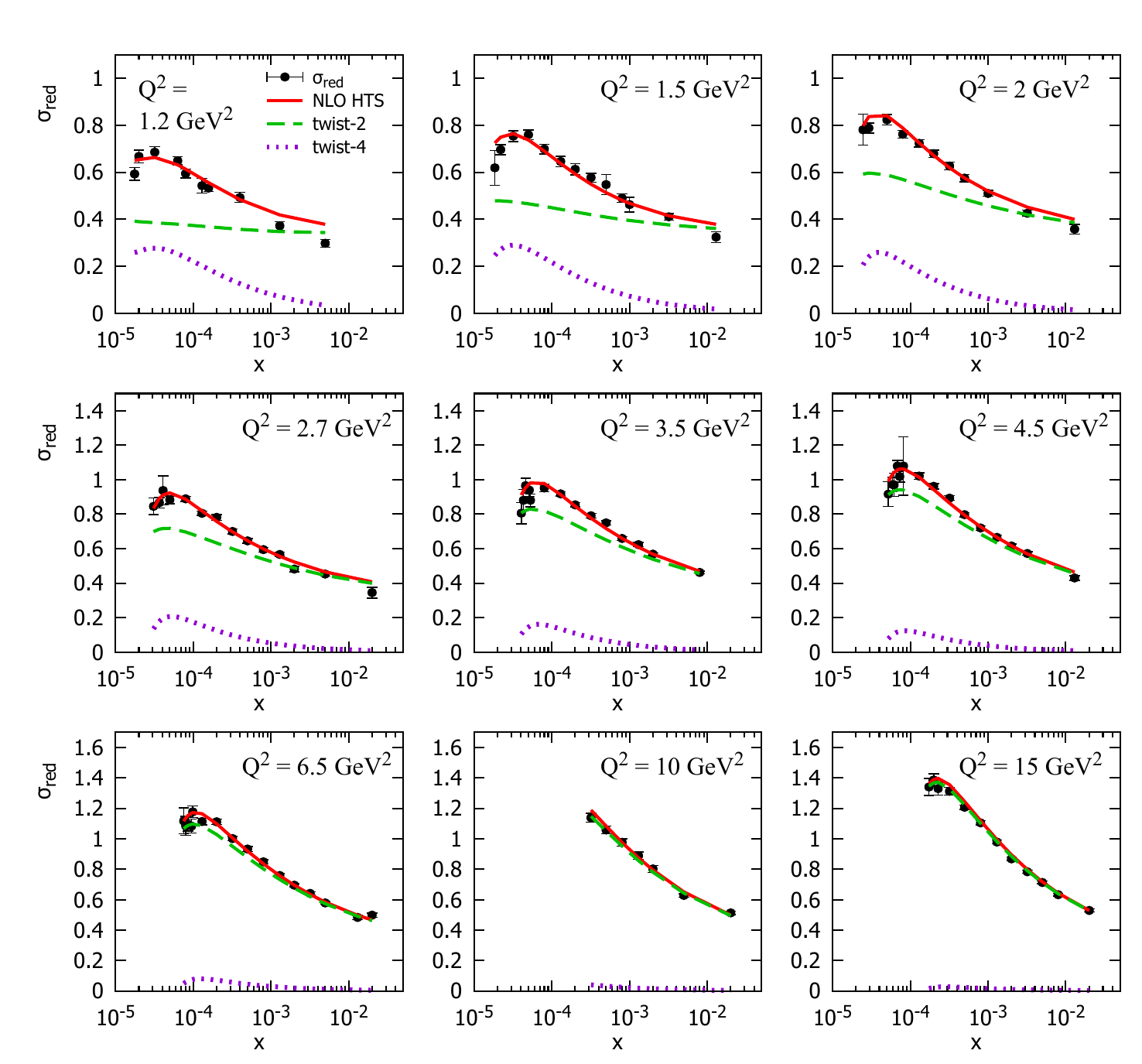}}
\caption{
The combined low-$Q^2$ HERA inclusive NC $e^+p$ reduced cross sections compared to the NLO HTS fits (full red line).
Also shown twist~2 (dashed green line) and twist~4 (dotted purple line) contributions.}
\label{fig:sigred_NLO}
\end{figure}

In Fig.\ \ref{fig:sigred_NLO} the data for the reduced cross sections for $1.2\;\mathrm{GeV}^2 < Q^2 < 15\;\mathrm{GeV}^2$ are compared to the HTS fit results. In order to illustrate the twist content of the proton structure function we also show the twist decomposition of the reduced cross section at the NLO level. The twist~4 contribution makes up to 75$\%$ of the twist~2 contribution at $Q^2=1.2\GeV^2$ and $x=3\cdot 10^{-5}$, where the higher-twist effects are largest. In this kinematic region the twist~4 effects are estimated to provide about 40$\%$ of the reduced cross section. As expected, the higher-twist contribution is suppressed with increasing $x$, and at $x=3\cdot 10^{-4}$ the twist~4 correction is reduced to about 20$\%$ of the total value. The relative importance of the higher-twist correction decreases also with growing $Q^2$. Indeed, at $x=3\cdot 10^{-5}$ the relative twist~4 contribution is about 30\% at $Q^2=2\GeV^2$ and about 20$\%$ at $Q^2=3.5\GeV^2$. In the \fitName{HTS} fit at NLO, the relative higher-twist effect is below 10\% at $Q^2=6.5\GeV^2$.

\begin{figure}[h]
\centerline{\includegraphics[width=\szer]{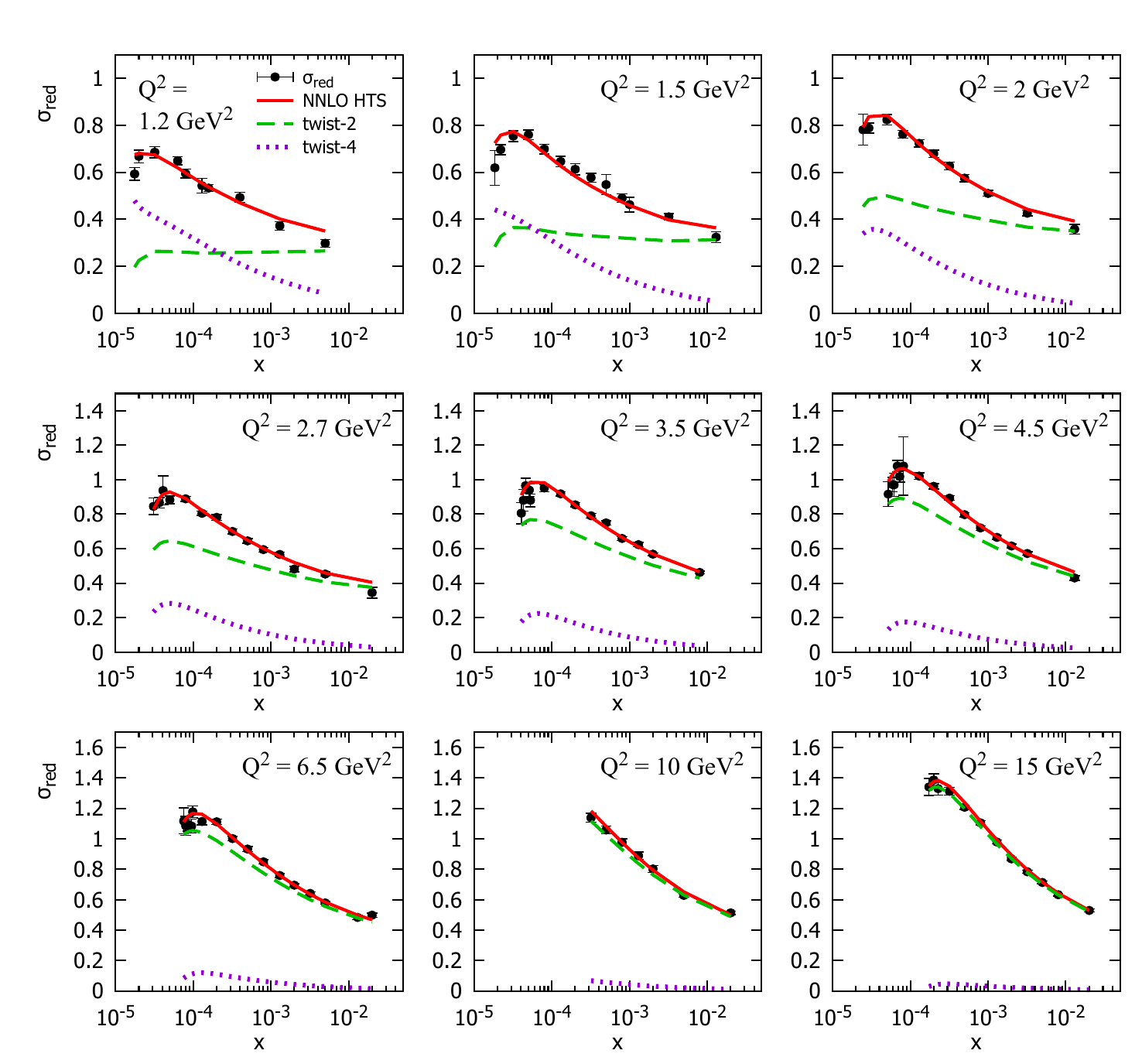}}
\caption{
The combined low-$Q^2$ HERA inclusive NC $e^+p$ reduced cross sections compared to the NLO HTS fits (full red line).
Also shown twist~2 (dashed green line) and twist~4 (dotted purple line) contributions.}
\label{fig:sigred_NNLO}
\end{figure}

The results of a similar investigation at NNLO are displayed in Fig.\ \ref{fig:sigred_NNLO} for $Q^2$ up to 15~GeV$^2$. In the NNLO DGLAP fits the higher-twist effects are found to be significantly stronger than in the NLO fits over the whole probed range of $Q^2$. In particular, in the NNLO fit at $Q^2=1.2\GeV^2$, the twist~4 correction is found to be larger than the leading-twist contribution for $x< 2\cdot 10^{-4}$, and the relative correction further grows towards small $x$, to reach about 200\% of the leading-twist term at the lowest available $x \simeq 2 \cdot 10^{-5}$. At $Q^2=2\GeV^2$ and the smallest~$x$, the twist~4 correction reaches about 80$\%$ of the twist~2 contribution, and at $Q^2=3.5\GeV^2$ the higher-twist term is still around 25$\%$ of the leading-twist term. Finally, the higher-twist correction reaches $\sim 10\%$ level at $Q^2=6.5\GeV^2$ and quickly decreases for larger $Q^2$.

The characteristic behaviour of the data at moderate $Q^2$ is a turn-over at small~$x$. This feature is not reproduced by the DGLAP fits without higher-twist corrections \cite{Abramowicz:2015mha} and the inclusion of higher-twist effects is necessary to provide a good description of this behaviour \cite{Harland-Lang:2016yfn,Abt:2016vjh}. Hence the turn-over may be considered to be a signature of the higher-twist contributions. The HTS fits reproduce well this shape both at NLO and NNLO.

\begin{figure}[htb]
\centerline{
\includegraphics[width=\szer]{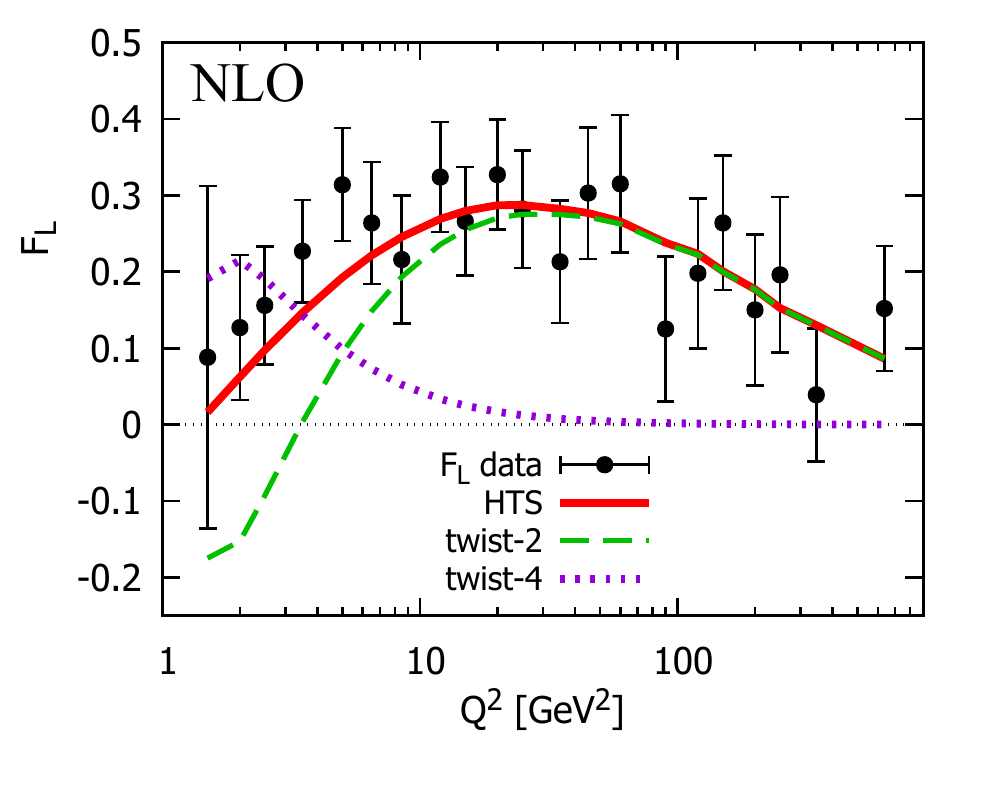}
\includegraphics[width=\szer]{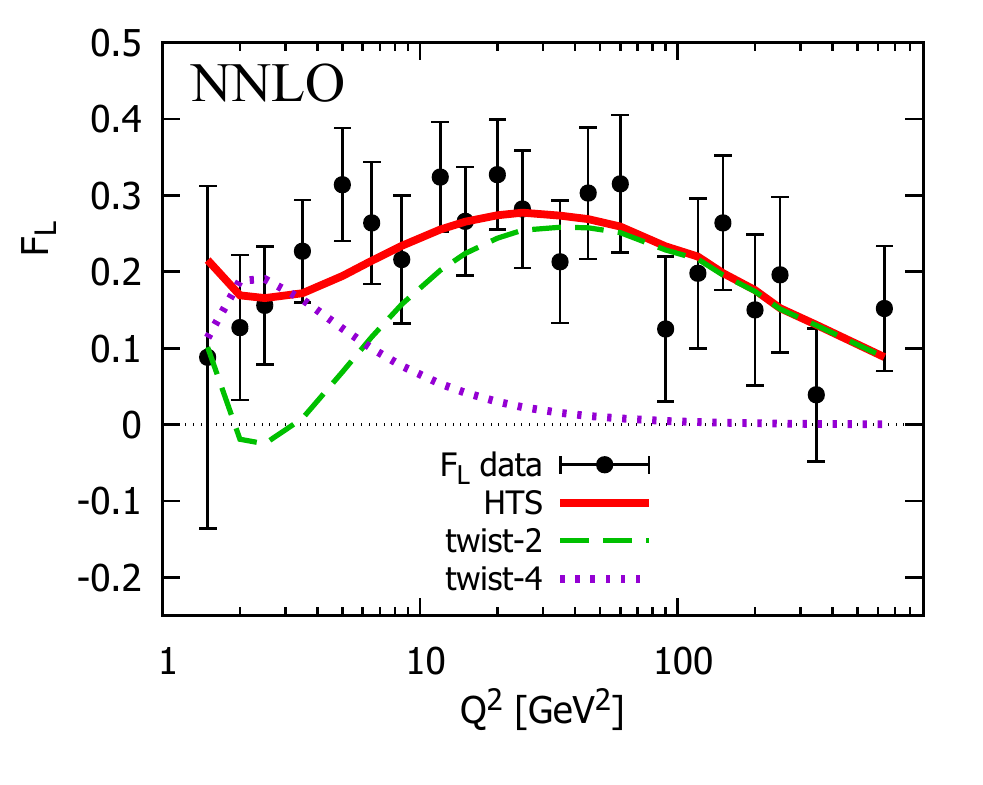}
}
\caption{The predictions for \FL from the \fitName{HTS} fit compared to the H1 data \cite{Andreev:2013vha}.
Also shown are twist~2 (dashed green line) and twist~4 (dotted purple line) contributions.}
\label{fig:FL}
\end{figure}

In the existing analyses of the combined HERA data with higher-twist corrections \cite{Abt:2016vjh} a satisfactory description of the $\FL$ data at smaller $Q^2$ has not been achieved within the NNLO framework \cite{Abt:2016vjh}. The predictions for $\FL$ obtained in our approach are shown in Fig.\ \ref{fig:FL} in comparison to the experimental data from H1 \cite{Andreev:2013vha}. The $\FL$ data are well described down to $Q^2 = 1$~GeV$^2$. Note that the plotted $\FL$ data were not directly fitted, the $\FL$ contribution was treated in the fits only as a part of the reduced cross section $\sigma_{\mathrm{red}}$. The higher-twist contributions are found to be important in $\FL$ at small and moderate $Q^2$. In particular, the twist~4 term dominates for $Q^2 < 5$ GeV$^2$ ($Q^2 < 6$ GeV$^2$) for the NLO fit (NNLO fit). 
Remarkably,  in the NNLO fit, the higher-twist contribution is still visible at 10\% level up to a sizeable scale of $Q^2 \simeq 20\GeV^2$. This shows that the longitudinal structure function is particularly sensitive to the higher-twist effects and it may be used as their effective probe.

\szer=0.5\columnwidth


\szer=0.8\columnwidth

\begin{figure}[htb]
\centerline{\includegraphics[width=\szer]{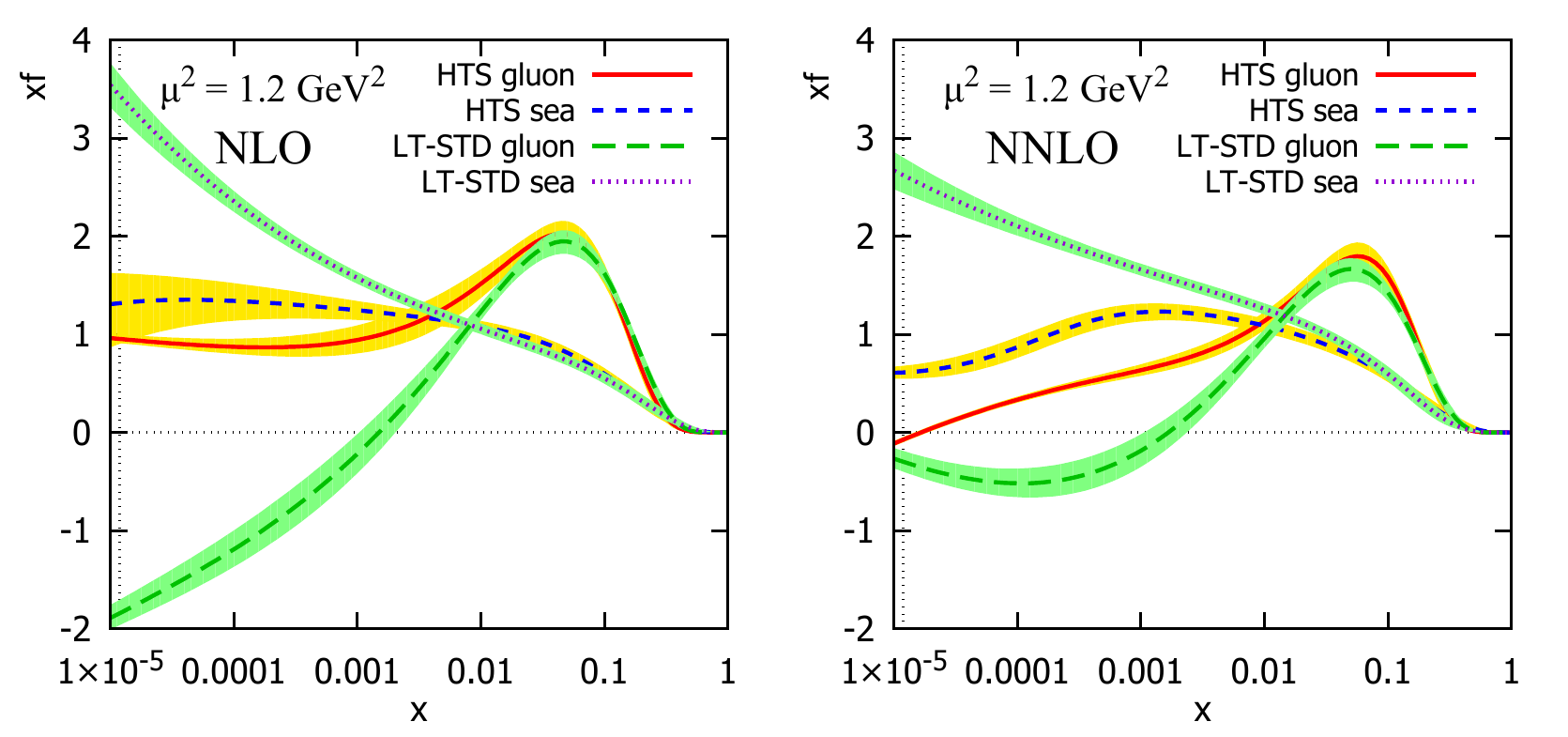}}
\centerline{\includegraphics[width=\szer]{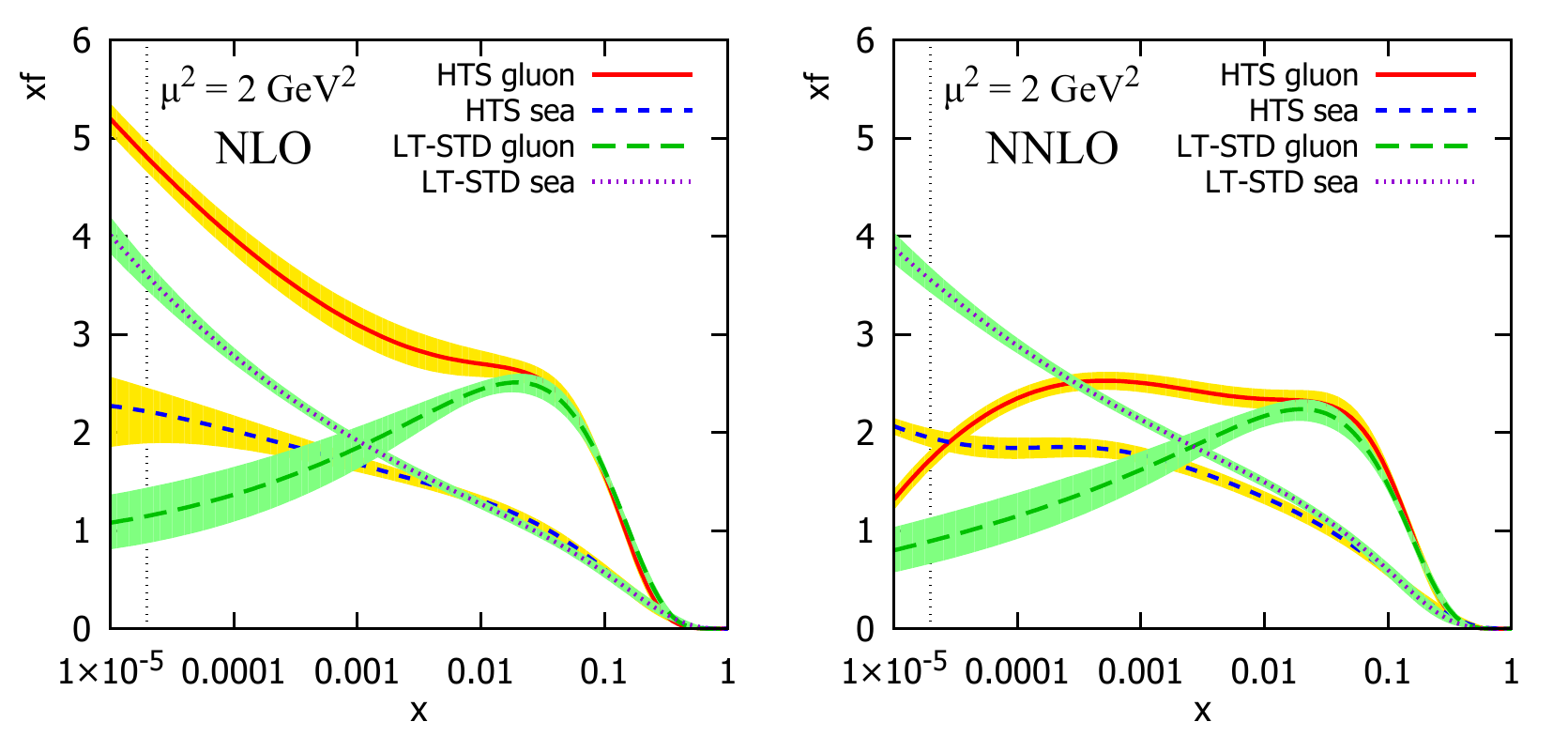}}
\centerline{\includegraphics[width=\szer]{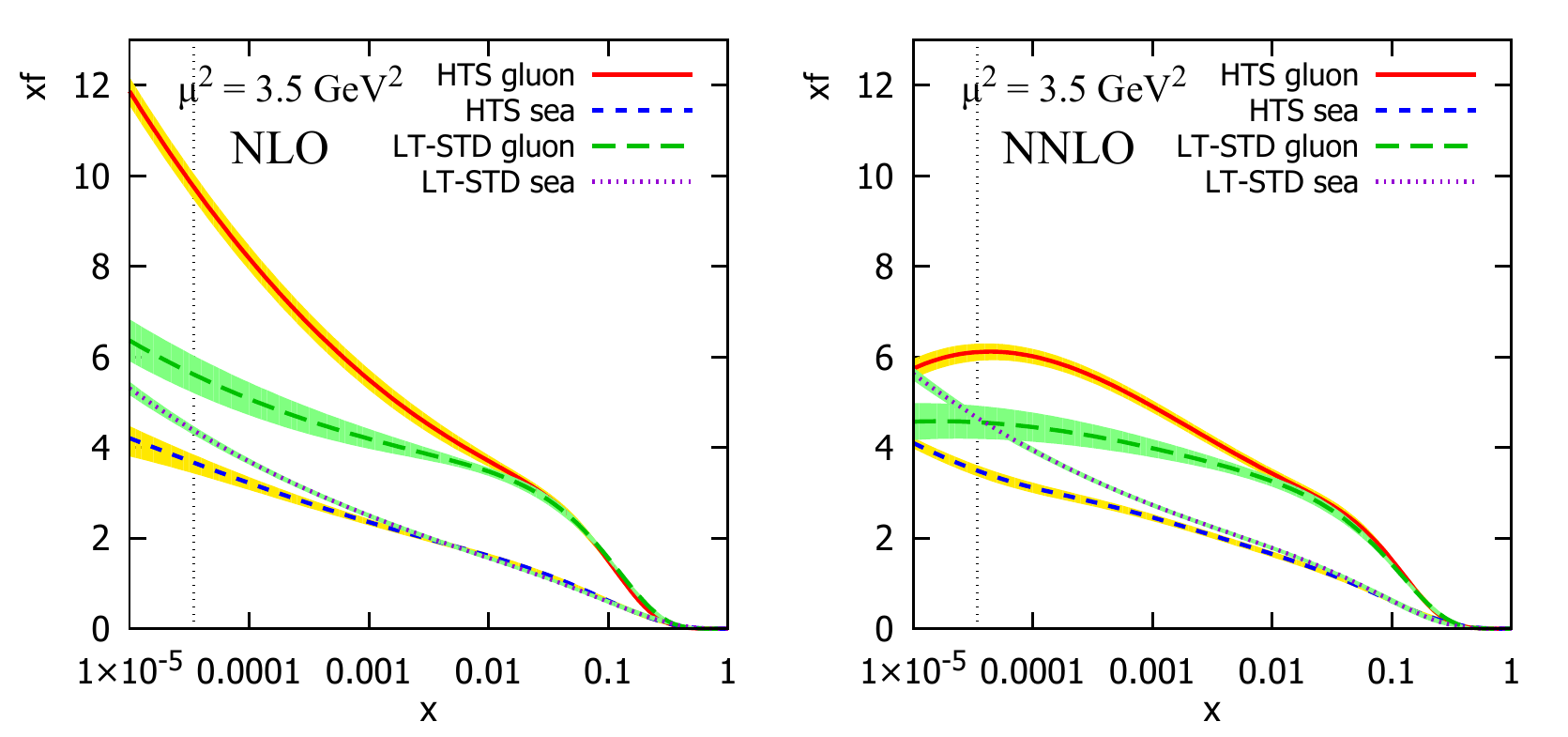}}
\caption{Comparison of the \fitName{HTS} and \fitName{LT-STD} PDFs at $\mu^2 = 1.2\GeV^2$ (top),  $\mu^2 = 2.0\GeV^2$ (middle) and $\mu^2 = 3.5\GeV^2$ (bottom). For $\mu^2 = 1.2\GeV^2$ the \fitName{LT-STD} PDFs are extrapolated below the parameterisation range. The experimental uncertainties are shown.}
\label{fig:pdfs-1p2}
\end{figure}

\szer=0.9\columnwidth

\begin{figure}[htb]
\centerline{\includegraphics[width=\szer]{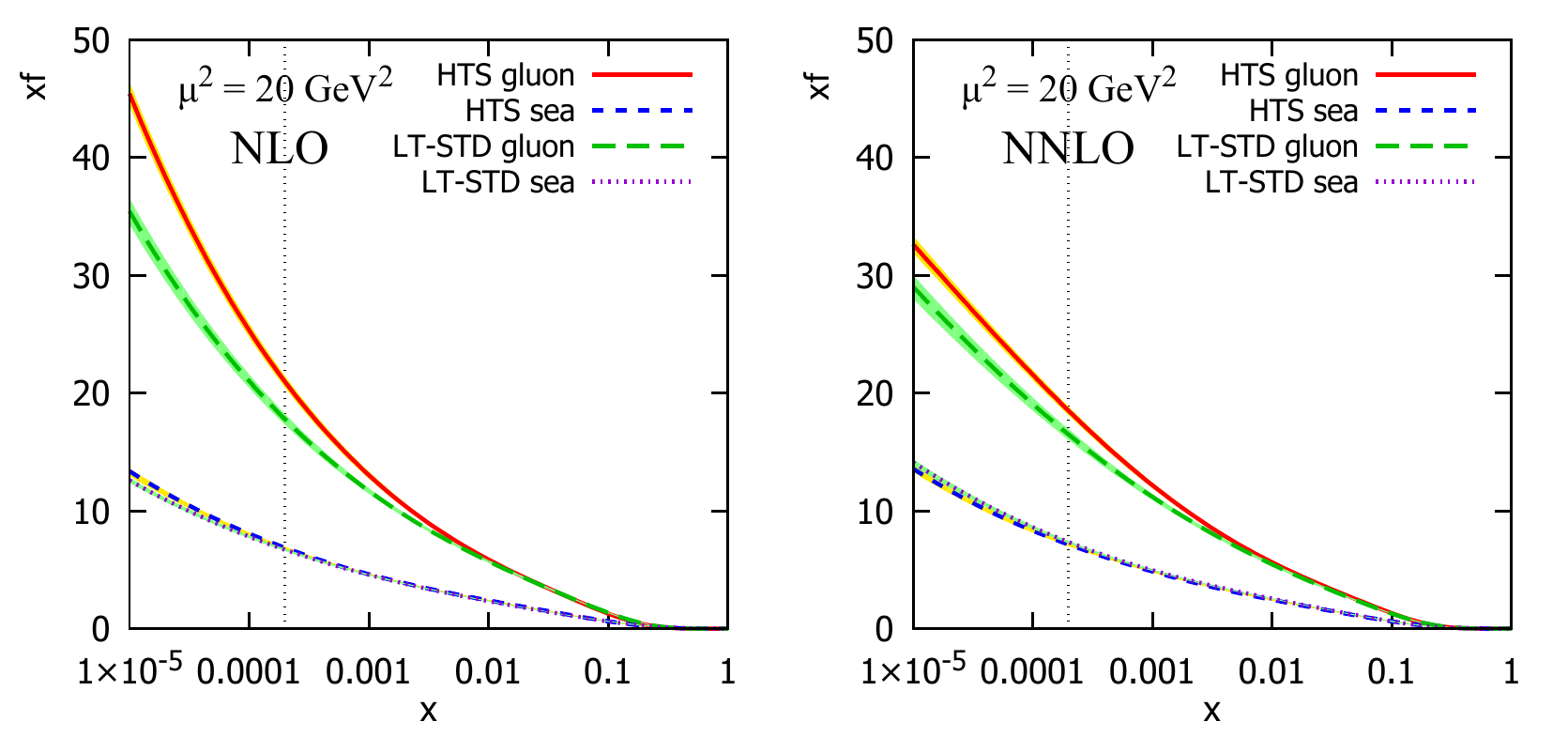}}
\centerline{\includegraphics[width=\szer]{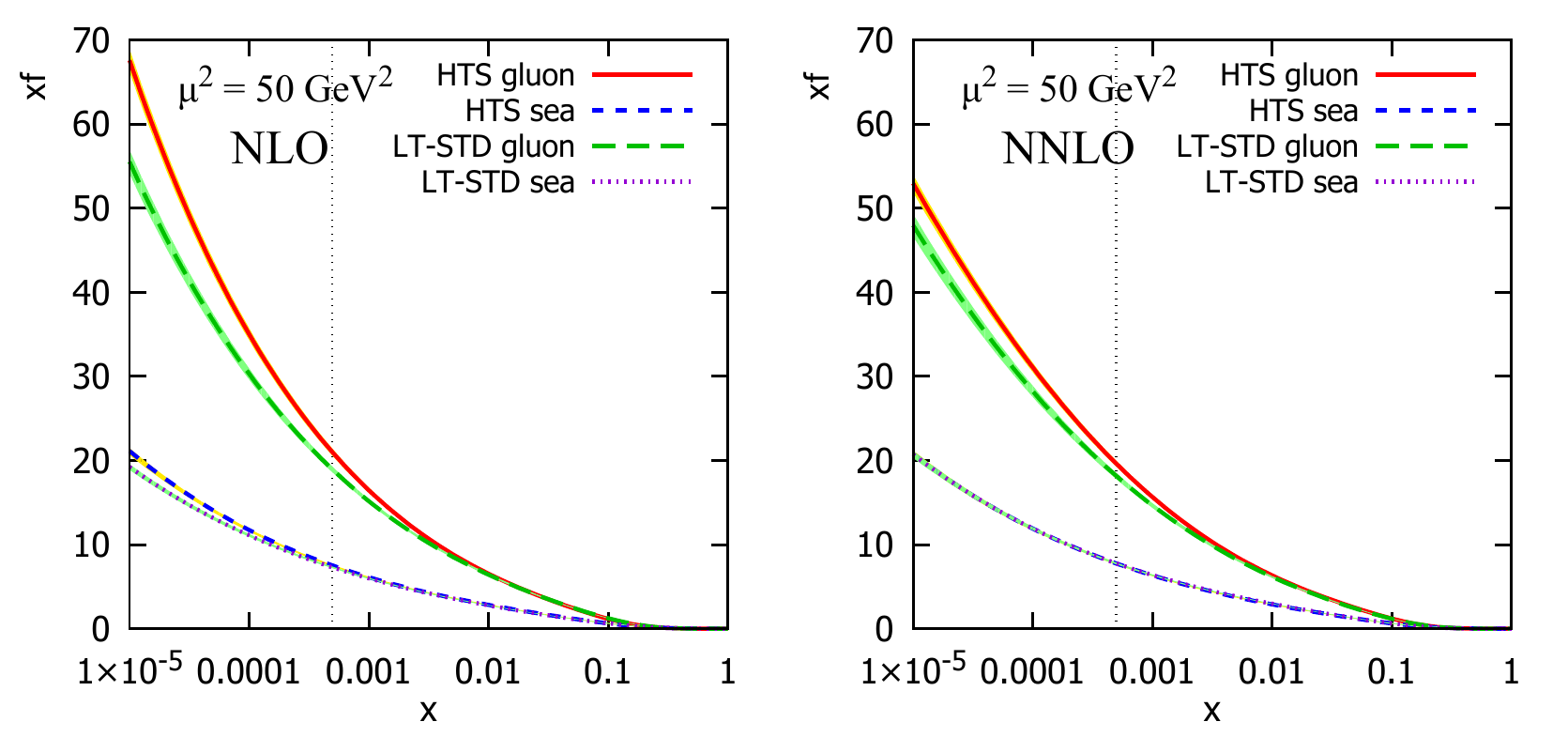}}
\caption{Comparison of the \fitName{HTS} and \fitName{LT-STD} PDFs at $\mu^2 = 20\GeV^2$ (top) and $\mu^2 = 50\GeV^2$ (bottom). The experimental uncertainties are shown.}
\label{fig:pdfs-20}
\end{figure}

\szer=0.5\columnwidth
\begin{figure}[htb]
\centerline{
  \includegraphics[width=\szer]{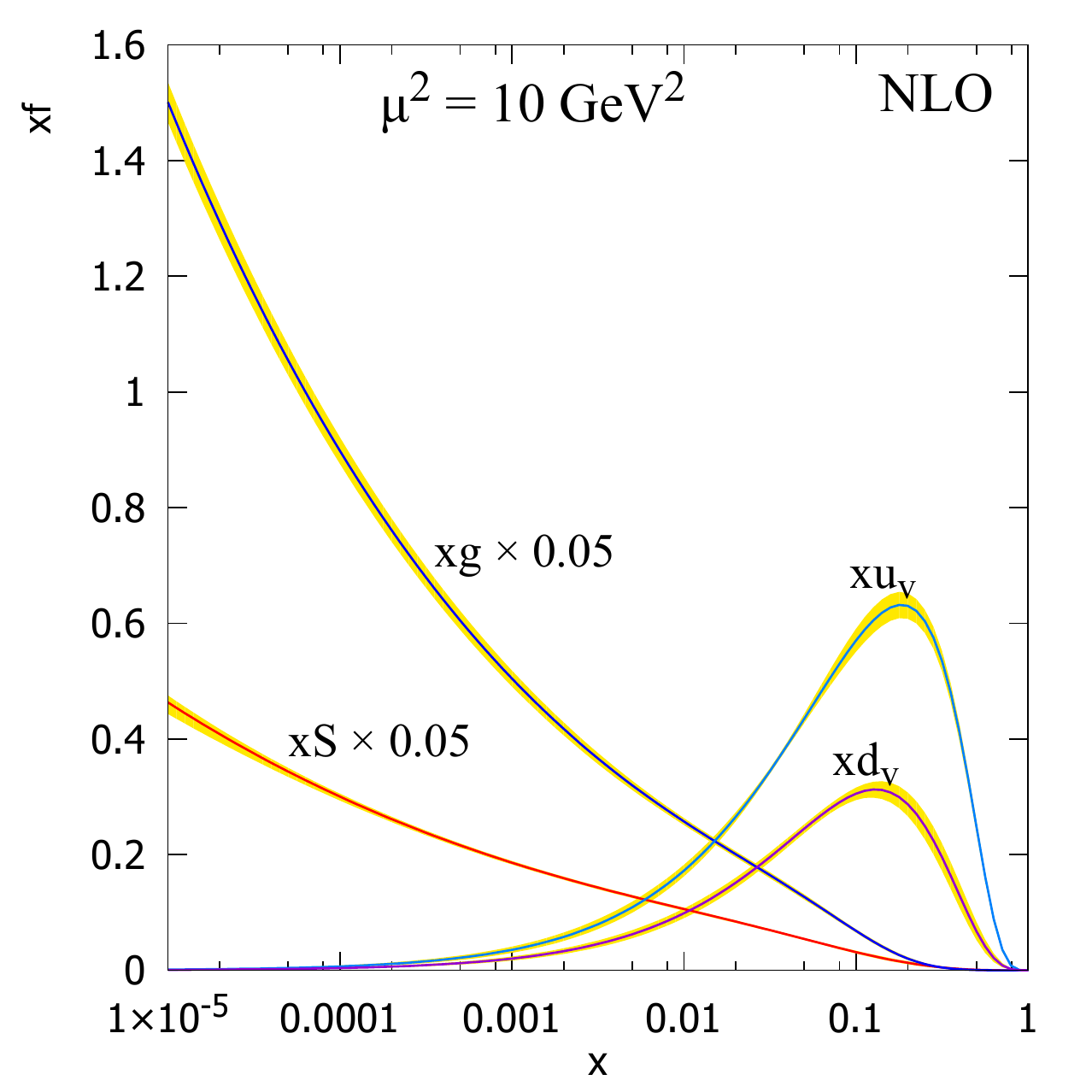}
  \includegraphics[width=\szer]{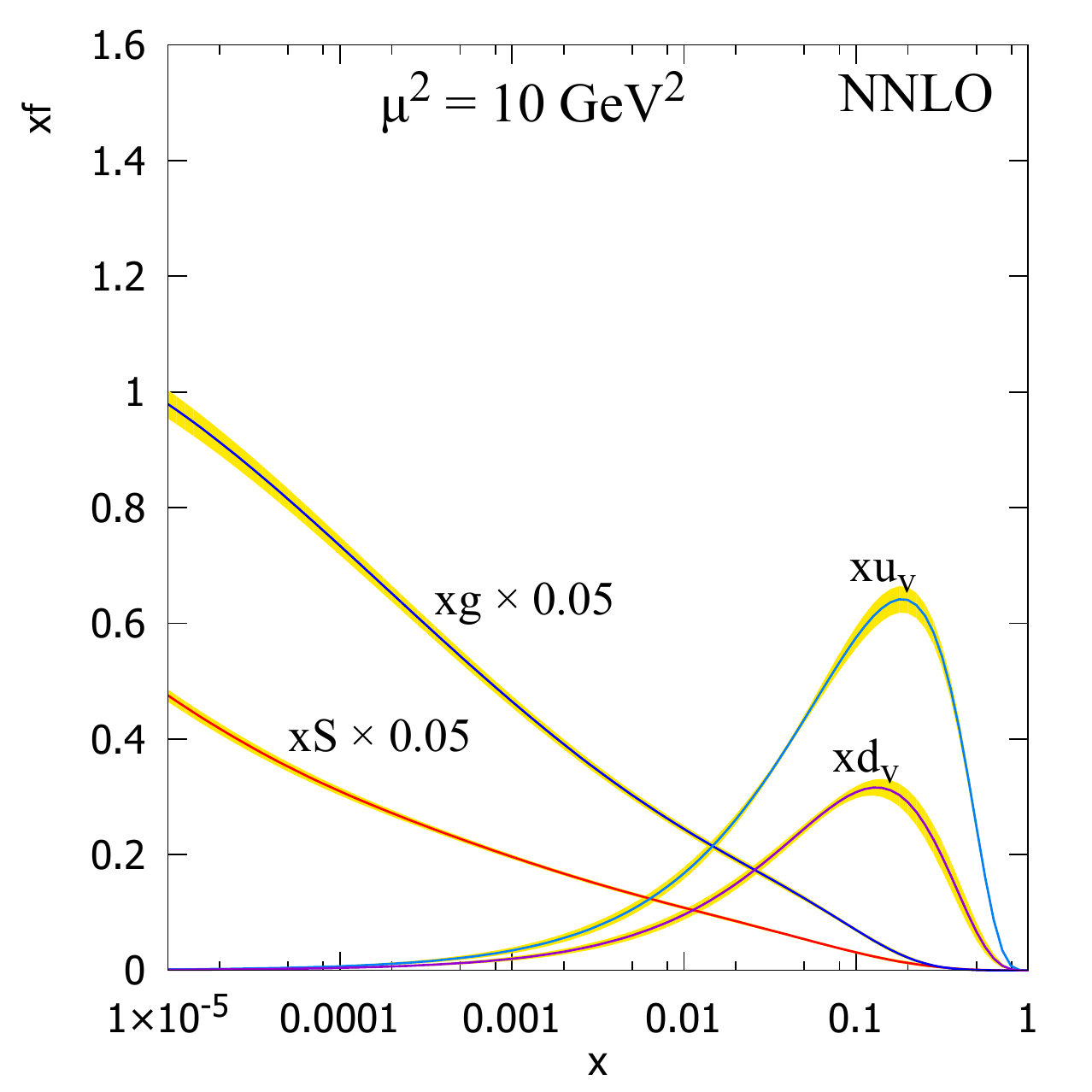}
}
\caption{The PDFs at $\mu^2 = 10\GeV^2$ obtained from the \fitName{HTS} fit. The experimental uncertainties are shown.}
\label{fig:HTS-pdfs-10}
\end{figure}

\subsection{Impact of the higher-twist effects on the PDFs}

One of the key goals of the present analysis is to understand the impact of the higher-twist corrections on the parton density functions. It is expected that inclusion of the higher-twist corrections alters the resulting PDFs, hence affecting the predictions in which the PDFs are used. Below we present the PDFs obtained from the \fitName{HTS} fits (including the higher-twist effects) compared to the PDFs obtained within the standard framework, which we denote as the \fitName{LT-STD} fit. The determined PDFs are presented with the corresponding experimental uncertainties. The included higher-twist corrections are larger in the small~$x$ region, so they affect mostly the gluon and sea PDFs at small~$x$. Hence, in the figures we show only these parton distributions. We have checked that the effect of the higher-twist corrections on the valence quark distribution is much smaller than for the gluon and sea PDFs.

In Fig.\ \ref{fig:pdfs-1p2} we compare the gluon and sea PDFs at small scales, $\mu^2 = 1.2$, 2.0 and 3.5 GeV$^2$, with (\fitName{HTS}) and without (\fitName{LT-STD}) the higher-twist corrections, at the NLO and NNLO accuracy. The input condition for the \fitName{HTS} fit is assumed at  $\mu_0^2 = 1$~GeV$^2$ and it includes the saturation damping effect in the gluon and sea PDFs. The standard fit (\fitName{LT-STD}) starting scale is $\mu^2_0 = 1.9\GeV^2$, so for $\mu^2 = 1.2$~GeV$^2$ the \fitName{LT-STD} PDFs should be treated as extrapolations only. 
The dotted vertical lines in the plots mark the minimal kinematically allowed value of $x$ in HERA measurements. Hence the PDFs values to the left of this line are also extrapolations from the region of available data towards smaller~$x$.  

It is clear from Fig.\ \ref{fig:pdfs-1p2} that inclusion of the higher-twist and saturation effect leads to sizeable changes of the gluon and sea PDFs at small $x <0.01$ and small scales $\mu^2 < 3.5$~GeV$^2$ both at the NLO and NNLO accuracy. The effects found are larger for the NLO fits, where e.g.\ for $\mu^2 = 3.5$~GeV$^2$ and $x = 3\cdot 10^{-5}$ the \fitName{HTS} gluon is larger than the \fitName{LT-STD} gluon by about 60\% and the \fitName{HTS} sea PDF is smaller by more than 10\% than the \fitName{LT-STD} one. In the NNLO fits the differences between the gluon PDFs with and without the higher-twist effects tend to be smaller than in NLO fits, while the effect is larger in the sea quarks at NNLO. In general, at the smaller scales, the inclusion of the higher-twist effects leads to a larger gluon PDF and the reduced sea PDF. The differences found are not only in the shape and values of the PDFs but also in the relative behaviour of the gluon and sea distributions. In the standard approach those two are decoupled at the input scale whereas in the \fitName{HTS} approach the sea distribution at small~$x$ follows the gluon distribution already at the input scale.

In a similar way, \Fig{fig:pdfs-20} displays the impact of higher-twist effects on the PDFs at larger scales, $\mu^2 = 20$ and 50~GeV$^2$. The higher-twist effects are still significant in the gluon distribution at small~$x$. As for the smaller scales, the inclusion of the higher-twist effects leads to a larger gluon PDF. The change of the gluon PDF at the small~$x$ values corresponding to the kinematic lower limit at HERA is about 10\% for $\mu^2 = 20\GeV^2$, and  slightly below 10\% for $\mu^2 = 50$ GeV$^2$. At the larger scales the sea distribution is not  significantly affected by the higher-twist corrections. Note finally that the higher-twist corrections lead to larger changes in the PDFs for the NLO DGLAP framework than for the NNLO DGLAP one.

Fig.\ \ref{fig:HTS-pdfs-10} shows the independent PDFs at $\mu^2 = 10$~GeV$^2$ obtained from our best fits with higher-twist corrections and the parton saturation effects. The valence up and down quark, the sea and the gluon PDFs are shown with their experimental uncertainties at the NLO and NNLO accuracy. 


%% file: Disc.tex
\section{Discussion}
\label{Sec:Disc}

Let us recall the main results of this paper that substantiate the evidence of the significant higher-twist contributions in the inclusive DIS at HERA at small~$x$ and $Q^2$. The strongest point is the comparison of the fits to the combined HERA data on the reduced cross sections based on the DGLAP evolution with and without twist~4 corrections. For the data sample analysed, with $Q^2>1$~GeV$^2$, the inclusion of the twist~4 terms improves the $\chi^2/ $d.o.f.\ of the NLO fit from about 1.35 (for the pure DGLAP) to about 1.22. With about 1200 degrees of freedom of the fit, the statistical significance of the $\chi^2$ change corresponds to an improvement of the $p$-value by more than seven orders of magnitude. For the NNLO DGLAP fit we find an improvement of the $\chi^2/ $d.o.f.\ from about 1.5 to less than 1.2, so the statistical significance of the improvement is greater by many orders of magnitude that in the NLO DGLAP fits. On the other hand, the values of the $\chi^2/ $d.o.f.\ about 1.2 found in the DGLAP fits with twist~4 corrections are still uncomfortably larger than 1.0. Here, however, being aware of difficulties in treatment of the systematic errors of the measurement, we take as the reference level the value $\chi^2/ $d.o.f.\ = 1.15 that is obtained in all the fits for $\QQmin >10$~GeV$^2$, where the higher-twist corrections should be negligible. Given the crudeness of the higher-twist model applied, we find the change of the $\chi^2/ $d.o.f.\ from the reference level 1.15 to less than 1.2 in the best NNLO fit with higher twists to be sufficiently small to accept the model.

An issue that should be raised is a possible alternative explanation of the data by a pure DGLAP fit with a more flexible input parameterisation. Taking into account this possibility we studied in detail the effect of variation of the sea and gluon input PDFs at small~$x$ due to the inclusion of the fitted saturation damping effects. We found that these variations of the input PDFs do not change the conclusion on the presence of the higher-twist contributions. Moreover, the contribution of the higher twists found is characterised by a strong $Q^2$ dependence which cannot be mimicked by the features of the input parameterisations that are given at a fixed scale $\mu_0^2$. The explicit analysis of the $Q^2$ and $x$-dependencies of the higher-twist terms leading to the improvement of the fit quality shows full consistency with the assumption of the leading twist~4 exchange as the contribution necessary to well describe the data. In particular the obtained twist~4 correction is characterised by a leading $x^{-2\lambda}$ behaviour with $\lambda = 0.26$ for the NLO fit and  $\lambda = 0.35$ for the NNLO fit---the values strongly supporting the interpretation of the twist~4 terms as coming from the double hard gluon ladder exchange, as expected from the QCD analyses of the twist~4 evolution at small~$x$.

The higher twist interpretation advocated in this paper should be also confronted with recent interesting results obtained for the vector meson production \cite{Jones:2016ldq}, where power corrections to the coefficient functions in high energy scattering were derived, that do not come from the leading quasi-partonic 4-gluon twist~4 exchange{\footnote{We are grateful to Alan Martin for pointing out this to us.}. Although we agree that such corrections should also contribute to the DIS cross sections, we stress that the $x$-dependence of the twist~4 corrections ({\em i.e.} the power corrections in $1/Q^2$) found in the present analysis is consistent with the dominance of the exchange of two hard gluonic ladders.

The results presented here are compatible with the two recent papers \cite{Harland-Lang:2016yfn,Abt:2016vjh} in which a similar conclusion was reached that the combined HERA data require an inclusion of the twist~4 corrections at small~$x$ and $Q^2$. These two analyses have been performed using a similar method of the $\chi^2$~scan as a function of $\QQmin$ for DGLAP fits with and without the higher-twist corrections. In both approaches the data were selected with $Q^2 >2$~GeV$^2$ and no sensitivity was found to the higher-twist corrections in $\Ftwo$. Hence, in the central model of both the approaches, only $\FL$ receives the twist~4 correction in the form of $\Delta^{(\tau=4)} \FL (x,Q^2)= A \FL (x,Q^2) / Q^2$. The improvements and extensions of the present analysis are the following. The model of higher-twist corrections we propose here is based on general features of multiple scattering in QCD at high energies. In particular it includes the known results on the leading evolution of twist~4 quasi-partonic operators \cite{BFKL2,Bartels:1993it}. Therefore the model offers an explicit physics interpretation of the result and allows to obtain a deeper insight into the $x$-dependence of higher-twist corrections in both $\FL$ and $\FT$ (or $\Ftwo$). Furthermore, in order to enhance the sensitivity to the higher-twist effects, we extended the $Q^2$ range of the fitted data down to $Q^2=1$~GeV$^2$. Indeed, with this extension the fits are sensitive to twist~4 effects in both $\FL$ and $\Ftwo$ and the $x$-dependence of the twist~$4$ corrections is well constrained. Hence, the additional conclusion we find w.r.t.\ the previous studies is the steep $x^{-2\lambda}$ dependence of the twist~4 corrections, while the twist~4 correction $\Delta^{(\tau=4)} \FL (x,Q^2)$ of Refs.\ \cite{Harland-Lang:2016yfn,Abt:2016vjh} is characterised by a single gluonic ladder exchange $\sim x^{-\lambda}$. It is intriguing why the fits do not distinguish clearly between these two different scenarios. We propose the following interpretation of this apparent discrepancy. As follows from the performed twist decomposition of the reduced cross section in the best fits, the higher-twist corrections become significant in the bins of the lowest $Q^2$, below $Q^2 = 5$~GeV$^2$ and at the smallest available $x$. Therefore there is only a moderate region of the kinematic space (and a moderate number of the data points in this space) to pin down the $x$-dependence of the higher-twist corrections with a high precision, see a discussion of this issue performed in Ref.\ \cite{Harland-Lang:2016yfn}. In Refs.\ \cite{Harland-Lang:2016yfn,Abt:2016vjh} the data points were taken with $Q^2 > 2$~GeV$^2$ so their sensitivity to the $x$-dependence of the higher-twist corrections is smaller than in the present analysis. Furthermore in the present approach we introduce additional freedom of the sea input parameterisation at small~$x$ motivated by parton saturation damping effects. This feature leads to a reduced sea PDF at small $x$ and small scales w.r.t.\ the standard sea PDF and a modified physics picture emerges. In \cite{Harland-Lang:2016yfn,Abt:2016vjh} the sizeable higher-twist corrections are found in $\FL$ and they are negligible in $\Ftwo$. In the \fitName{HTS} NNLO fit obtained in the present analysis we find the sizeable positive higher-twist corrections in both $\FL$ and $\Ftwo$ at small~$x$ and $Q^2$, and certain suppression of the leading twist contribution to $\Ftwo$ due to the saturation damping of the sea. 
Finally the model of higher-twist correction of Ref.\ \cite{Abt:2016vjh} is simple and efficient, however it leads to unphysical behaviour of the predicted $\FL$ at small $Q^2$ at NNLO. In summary, our results are consistent with the main findings of Refs.\ \cite{Harland-Lang:2016yfn,Abt:2016vjh} but we offer a more flexible approach to the higher-twist corrections with a clear physical interpretation. The proposed framework can be successfully extended down to $Q^2=1$~GeV$^2$ and gives a good description of $\FL$ over the whole $Q^2$ range.

A comment is in order on the fact that the sign of the obtained twist~4 correction to $\FL$ is positive and it disagrees with the predictions~\cite{BGBP,BGBM} from the Golec-Biernat--W\"{u}sthoff saturation model~\cite{GBW}. The result is also not intuitive in a simplified picture of the multiple scattering, in which the longitudinal virtual photon--proton scattering total cross section receives a positive correction from double scattering, while one could expect a negative shadowing correction. The sign of the twist~4 correction depends however, on the sign of the corresponding coefficient function. In the GBW model the twist~4 coefficient function is implicitly computed assuming eikonal coupling of two gluonic ladders to the virtual quark loop coming from the virtual photon fluctuation. Such a coupling is not what is predicted by QCD at small~$x$. The $\gamma^*$ coupling to four-gluon exchange was analysed in the leading logarithmic $1/x$ approximation in the framework of extended generalised leading logarithmic approximation \cite{Bartels:1992ym,Bartels:1994jj,EGGLA,Motyka:2014jpa} and it was found that the coupling occurs via triple pomeron vertex, which is distinctly different from the eikonal coupling. A more detailed analysis of the four-gluon coupling to the virtual photon was performed in Ref.\ \cite{Bartels:1999xt}. The definite prediction of the sign of the twist~4 corrections to the proton structure functions in this approach is not available yet, so it is interesting to obtain it and confront with the results of the fit.

Finally, let us discuss shortly the concept of the gluon and sea saturation damping effects in the input PDFs at small~$x$. The motivation to propose this damping for the gluon PDF comes from the properties of the solution of the Balitsky--Kovchegov equation \cite{BK:Bal,BK:Kov} and for the sea PDFs we additionally assume that the input sea PDF at small~$x$ is driven by the gluon PDF. The damping may be understood as a result of unitarity corrections due to the multiple scattering effects below the scale of the input parameterisation.
For the gluon PDF, the DGLAP fits (with or without higher-twist corrections) with the conventional parameterisation of the input gluon PDF, with the small~$x$ asymptotics $xf_g(x,\mu_0^2) \sim x^{B_g}$, yield a positive $B_g$ when $\mu_0^2$ is sufficiently small, inline with the saturation damping property. So, for the gluon PDF the saturation damping is implicit in the standard parameterisation. For the sea PDFs, however, the saturation damping leads to a significant modification of the input shape at $\mu_0^2 =1$~GeV$^2$. The saturation damping effect of the sea distribution is found to be supported by the data in the NNLO DGLAP fit with the higher-twist corrections. The inclusion of the saturation damping effects in this fit improves the $\chi^2/$d.o.f.\  significantly for moderate $Q^2$. In particular for the complete data sample the $\chi^2/$d.o.f.\ goes down from 1.25 to 1.2. In this fit, the characteristic value of $x$ for which the saturation damping effects turn on at $\mu_0^2 =1$~GeV$^2$ was found to be $\hat x \simeq 2\cdot 10^{-4}$, in consistence with the earlier analyses of the parton saturation at small~$x$ \cite{GBW,Kowalski:2006hc}.


%% file: dis-ht.final.bbl
\begin{thebibliography}{99}


\bibitem{Abramowicz:2015mha}
  H.~Abramowicz {\it et al.} [H1 and ZEUS Collaborations],
  {\it ``Combination of measurements of inclusive deep inelastic ${e^{\pm }p}$ scattering cross sections and QCD analysis of HERA data,''}
  Eur.\ Phys.\ J.\ C {\bf 75} (2015) no.12,  580.


\bibitem{Bartels:1993it}
  J.~Bartels and M.~G.~Ryskin,
  {\it ``The Analytic structure of the anomalous dimension of the four gluon operator in deep inelastic scattering,''}
  Z.\ Phys.\ C {\bf 62} (1994) 425.



\bibitem{BFKL2}
A.~P.~Bukhvostov, G.~V.~Frolov, L.~N.~Lipatov and E.~A.~Kuraev,
{\it ``Evolution Equations for Quasi-Partonic Operators,''}
Nucl.\ Phys.\ B {\bf 258} (1985) 601.



\bibitem{Bartels:1993ke}
  J.~Bartels and M.~G.~Ryskin,
{\it ``Absorptive corrections to structure functions at small $x$,''}
  Z.\ Phys.\ C {\bf 60} (1993) 751.

\bibitem{Martin:1998kka}
  A.~D.~Martin and M.~G.~Ryskin,
  {\it ``Higher twists in deep inelastic scattering,''}
  Phys.\ Lett.\ B {\bf 431} (1998) 395.

\bibitem{Martin:1998np}
  A.~D.~Martin, R.~G.~Roberts, W.~J.~Stirling and R.~S.~Thorne,
  {\it ``Scheme dependence, leading order and higher twist studies of MRST partons,''}
  Phys.\ Lett.\ B {\bf 443} (1998) 301.


\bibitem{MoSadSlo}
L.~Motyka, M.~Sadzikowski and W.~S\l{}ominski,
{\it ``Evidence of strong higher twist effects in diffractive DIS at HERA at moderate $Q^2$,''}
Phys.\ Rev.\ D {\bf 86} (2012) 111501.


\bibitem{Harland-Lang:2016yfn}
  L.~A.~Harland-Lang, A.~D.~Martin, P.~Motylinski and R.~S.~Thorne,
  {\it ``The impact of the final HERA combined data on PDFs obtained from a global fit,''}
  Eur.\ Phys.\ J.\ C {\bf 76} (2016) no.4,  186.


\bibitem{Abt:2016vjh}
I.~Abt, A.~M.~Cooper-Sarkar, B.~Foster, V.~Myronenko, K.~Wichmann and M.~Wing,
{\it ``Study of HERA ep data at low Q$^2$ and low $x_{\mathrm{Bj}}$ and the need for higher-twist corrections to standard perturbative QCD fits,''}
Phys.\ Rev.\ D {\bf 94} (2016) no.3, 034032.

\bibitem{Collaboration:2010ry}
  F.~D.~Aaron {\it et al.} [H1 Collaboration],
 {\it ``Measurement of the Inclusive $e^{\pm}p$ Scattering Cross Section at High Inelasticity $y$ and of the Structure Function $F_L$,''}
  Eur.\ Phys.\ J.\ C {\bf 71} (2011) 1579.

\bibitem{Andreev:2013vha}
  V.~Andreev {\it et al.} [H1 Collaboration],
{\it ``Measurement of inclusive $e p$ cross sections at high $Q^2$ at $\sqrt s =$ 225 and 252 GeV and of the longitudinal proton structure function $F_L$ at HERA,''}
  Eur.\ Phys.\ J.\ C {\bf 74} (2014) no.4,  2814.

\bibitem{Abramowicz:2014jak}
  H.~Abramowicz {\it et al.} [ZEUS Collaboration],
  {\it ``Deep inelastic cross-section measurements at large $y$ with the ZEUS detector at HERA,''}
  Phys.\ Rev.\ D {\bf 90} (2014) no.7,  072002.






\bibitem{BGBP}
J.~Bartels, K.~J.~Golec-Biernat and K.~Peters,
{\it ``An Estimate of higher twist at small $x_{\mathrm{B}}$ and low $Q^2$ based upon a saturation model,''}
Eur.\ Phys.\ J.\ C {\bf 17} (2000) 121.

\bibitem{BGBM}
J.~Bartels, K.~Golec-Biernat and L.~Motyka,
{\it ``Twist expansion of the nucleon structure functions, $F_2$ and $F_L$, in the DGLAP improved saturation model,''}
Phys.\ Rev.\ D {\bf 81} (2010) 054017.



\bibitem{GBW}
K.~J.~Golec-Biernat and M.~W\"{u}sthoff,
{\it ``Saturation effects in deep inelastic scattering at low $Q^2$ and its implications on diffraction,''}
Phys.\ Rev.\ D {\bf 59} (1998) 014017;
%
{\it ``Saturation in diffractive deep inelastic scattering,''}
Phys.\ Rev.\ D {\bf 60} (1999) 114023.

\bibitem{Alekhin:2014irh}
  S.~Alekhin {\it et al.},
  {\it ``HERAFitter,''}
  Eur.\ Phys.\ J.\ C {\bf 75} (2015) no.7,  304.


\bibitem{Kowalski:2006hc}
  H.~Kowalski, L.~Motyka and G.~Watt,
  {\it ``Exclusive diffractive processes at HERA within the dipole picture,''}
  Phys.\ Rev.\ D {\bf 74} (2006) 074016.





\bibitem{BraunTwist}
V.~M.~Braun, A.~N.~Manashov and J.~Rohrwild,
{\it ``Renormalization of Twist-Four Operators in QCD,''}
Nucl.\ Phys.\ B {\bf 826} (2010) 235.






\bibitem{Bartels:1994jj}
  J.~Bartels and M.~W\"{u}sthoff,
  {\it ``The Triple Regge limit of diffractive dissociation in deep inelastic scattering,''}
  Z.\ Phys.\ C {\bf 66} (1995) 157.


\bibitem{BK:Bal}
  I.~Balitsky,
  {\it ``Operator expansion for high-energy scattering,''}
  Nucl.\ Phys.\ B {\bf 463} (1996) 99.


\bibitem{BK:Kov}
  Y.~V.~Kovchegov,
  {\it ``Small $x$ $F_2$ structure function of a nucleus including multiple pomeron exchanges,''}
  Phys.\ Rev.\ D {\bf 60} (1999) 034008;
%
  {\it ``Unitarization of the BFKL pomeron on a nucleus,''}
  Phys.\ Rev.\ D {\bf 61} (2000) 074018.


\bibitem{Bartels:1992ym}
  J.~Bartels,
  {\it ``Unitarity corrections to the Lipatov pomeron and the small $x$ region in deep inelastic scattering in QCD,''}
  Phys.\ Lett.\ B {\bf 298} (1993) 204;
%
  {\it ``Unitarity corrections to the Lipatov pomeron and the four gluon operator in deep inelastic scattering in QCD,''}
  Z.\ Phys.\ C {\bf 60} (1993) 471.


\bibitem{EGGLA}
  J.~Bartels and C.~Ewerz,
{\it ``Unitarity corrections in high-energy QCD,''}
  JHEP {\bf 9909} (1999) 026.




\bibitem{NZ}
N.~N.~Nikolaev and B.~G.~Zakharov,
{\it ``Color transparency and scaling properties of nuclear shadowing in deep inelastic scattering,''}
Z.\ Phys.\ C {\bf 49} (1991) 607.




\bibitem{Mueller:1993rr}
  A.~H.~Mueller,
  {\it ``Soft gluons in the infinite momentum wave function and the BFKL pomeron,''}
  Nucl.\ Phys.\ B {\bf 415} (1994) 373;
%
  A.~H.~Mueller and B.~Patel,
  {\it ``Single and double BFKL pomeron exchange and a dipole picture of high-energy hard processes,''}
  Nucl.\ Phys.\ B {\bf 425} (1994) 471.


\bibitem{Kutak:2003bd}
  K.~Kutak and J.~Kwieci\'{n}ski,
  {\it ``Screening effects in the ultrahigh-energy neutrino interactions,''}
  Eur.\ Phys.\ J.\ C {\bf 29} (2003) 521.


\bibitem{Kutak:2004ym}
  K.~Kutak and A.~M.~Sta\'{s}to,
  {\it ``Unintegrated gluon distribution from modified BK equation,''}
  Eur.\ Phys.\ J.\ C {\bf 41} (2005) 343.

\bibitem{Bartels:2006ea}
  J.~Bartels, S.~Bondarenko, K.~Kutak and L.~Motyka,
  {\it ``Exclusive Higgs boson production at the LHC: Hard rescattering corrections,''}
  Phys.\ Rev.\ D {\bf 73} (2006) 093004.

\bibitem{Bondarenko:2006ft}
  S.~Bondarenko and L.~Motyka,
  {\it ``Solving effective field theory of interacting QCD pomerons in the semi-classical approximation,''}
  Phys.\ Rev.\ D {\bf 75} (2007) 114015.

\bibitem{GBMS}
  K.~J.~Golec-Biernat, L.~Motyka and A.~M.~Sta\'{s}to,
  {\it ``Diffusion into infrared and unitarization of the BFKL pomeron,''}
  Phys.\ Rev.\ D {\bf 65} (2002) 074037.



\bibitem{Thorne:1997ga}
  R.~S.~Thorne and R.~G.~Roberts,
  {\it ``An Ordered analysis of heavy flavor production in deep inelastic scattering,''}
  Phys.\ Rev.\ D {\bf 57} (1998) 6871.

\bibitem{Thorne:2006qt}
  R.~S.~Thorne,
  {\it ``A Variable-flavor number scheme for NNLO,''}
  Phys.\ Rev.\ D {\bf 73} (2006) 054019.

\bibitem{Thorne:2012az}
  R.~S.~Thorne,
  {\it ``Effect of changes of variable flavor number scheme on parton distribution functions and predicted cross sections,''}
  Phys.\ Rev.\ D {\bf 86} (2012) 074017.




\bibitem{Martin:2009iq}
  A.~D.~Martin, W.~J.~Stirling, R.~S.~Thorne and G.~Watt,
  {\it ``Parton distributions for the LHC,''}
  Eur.\ Phys.\ J.\ C {\bf 63} (2009) 189.


\bibitem{Botje:2010ay}
  M.~Botje,
  {\it ``QCDNUM: Fast QCD Evolution and Convolution,''}
  Comput.\ Phys.\ Commun.\  {\bf 182} (2011) 490.



\bibitem{Jones:2016ldq}
  S.~P.~Jones, A.~D.~Martin, M.~G.~Ryskin and T.~Teubner,
  {\it ``The exclusive $J/\psi$ process at the LHC tamed to probe the low $x$ gluon,''}
  Eur.\ Phys.\ J.\ C {\bf 76} (2016) no.11,  633.




\bibitem{Motyka:2014jpa}
L.~Motyka and M.~Sadzikowski,
{\it ``Twist decomposition of proton structure from BFKL and BK amplitudes,''}
Acta Phys.\ Polon.\ B {\bf 45} (2014) 11, 2079.


\bibitem{Bartels:1999xt}
  J.~Bartels and C.~Bontus,
  {\it ``An Estimate of twist four contributions at small $x_{\mathrm{B}}$ and low $Q^2$,''}
  Phys.\ Rev.\ D {\bf 61} (2000) 034009;
  J.~Bartels, C.~Bontus and H.~Spiesberger,
  {\it ``Factorization of twist four gluon operator contributions,''}
  hep-ph/9908411.






\end{thebibliography}
